\documentclass[prd,onecolumn,11pt,floatfix,altaffilletter,superscriptaddress,preprintnumbers,showpacs,showkeys,tightenlines,nofootinbib]{revtex4-2}
\usepackage[a4paper, total={6.5in, 9.5in}]{geometry}

\pdfoutput=1
\pdfoutput=1
\usepackage[utf8]{inputenc}
\usepackage{graphicx,xcolor}
\usepackage{hyperref}
\usepackage[T1]{fontenc}
\usepackage{multirow}
\usepackage{mathtools}
\usepackage{csquotes}
\usepackage[english]{babel}
\usepackage{subcaption}
\usepackage{amsmath}
\usepackage{amssymb}
\usepackage{siunitx}
\usepackage{xspace}
\usepackage[capitalise]{cleveref}
\usepackage[shortlabels]{enumitem}
\usepackage{xpatch}
\makeatletter
\patchcmd{\@ssect@ltx}
    {\addcontentsline{toc}{#1}{\protect\numberline{}#8}}
    {}
    {}
    {}
\makeatother

\makeatletter
\renewcommand\@makecaption[2]{%
  \par
  \vskip\abovecaptionskip
  \begingroup
   \small\rmfamily
    \begingroup
     \samepage
     \flushing
     \let\footnote\@footnotemark@gobble
     \@make@capt@title{#1}{#2}\par
    \endgroup
  \endgroup
  \vskip\belowcaptionskip
}

\begin{document}

\title{Constructing precisely quasi-isodynamic magnetic fields}

\author{Alan Goodman}
\email{alan.goodman@ipp.mpg.de}
\affiliation{Max-Planck-Institut f{\"u}r Plasmaphysik, 
            D-17491 Greifswald, Germany}

\author{Katia Camacho Mata}
\affiliation{Max-Planck-Institut f{\"u}r Plasmaphysik, 
            D-17491 Greifswald, Germany}

\author{Sophia A Henneberg}
\affiliation{Max-Planck-Institut f{\"u}r Plasmaphysik, 
            D-17491 Greifswald, Germany}
            
\author{Rogerio Jorge}
\affiliation{Instituto de Plasmas e Fusão Nuclear, Instituto Superior Técnico, Universidade de Lisboa, 1049-001 Lisboa, Portugal}

\author{Matt Landreman}
\affiliation{Institute for Research in Electronics and Applied Physics, University of Maryland, 
            College Park, MD 20742, USA}

\author{Gabriel Plunk}
\affiliation{Max-Planck-Institut f{\"u}r Plasmaphysik, 
            D-17491 Greifswald, Germany}
            
\author{H\r{a}kan Smith}
\affiliation{Max-Planck-Institut f{\"u}r Plasmaphysik, 
            D-17491 Greifswald, Germany}

\author{Ralf Mackenbach}
\affiliation{Eindhoven University of Technology, 
            5612 AZ, Eindhoven, Netherlands}

\author{Per Helander}
\affiliation{Max-Planck-Institut f{\"u}r Plasmaphysik, 
            D-17491 Greifswald, Germany}
            
\date{\today}

%===========================================================================
\begin{abstract}

We present a novel method for numerically finding quasi-isodynamic stellarator magnetic fields with excellent fast-particle confinement and extremely small neoclassical transport. The method works particularly well in configurations with only one field period. 
We examine the properties of these newfound quasi-isodynamic configurations, including their bootstrap currents, particle confinement, and available energy for trapped-electron driven turbulence, as well as the degree to which they change when a finite pressure profile is added.
We finally discuss the differences between the magnetic axes of the optimized solutions and their respective initial conditions, and conclude with the prospects for future quasi-isodynamic optimization.

\end{abstract}
%===========================================================================

\maketitle

% Show TOC
\vspace{-0.5cm}
\renewcommand{\baselinestretch}{0.85}\normalsize
\tableofcontents
\renewcommand{\baselinestretch}{1.0}\normalsize

\hfill\break
%===========================================================================
\section{Introduction}
\label{sec:intro}
%===========================================================================
\subsection{Stellarators}
The stellarator is a class of magnetic-confinement fusion devices that uses a magnetic field to confine plasmas inside a toroidal vessel. In order to avoid prompt losses of fusion-produced alpha particles, and to avoid large neoclassical transport, the magnetic field must be able to confine collisionless orbits on relatively long time scales. This can be achieved in an ``omnigenous'' field, in which particles drift towards the edge of the plasma in equal measure as they drift inwards \cite{hall1975,helander2014,Cary:1997zzb}. While this property was once considered practically impossible to achieve in stellarators, recent efforts in stellarator design \cite{landreman2022} were able to find fields with nearly perfect omnigenity.

Stellarators have several compelling properties as fusion reactor candidates, the most important one being the absence of plasma disruptions. In comparison with tokamaks, the plasmas in stellarators are relatively stable to kink modes and tearing modes thanks to the fact that the net toroidal plasma current is usually very small \cite{Boozer2021}. It does not vanish entirely, even in the absence of active current drive, due to the bootstrap current arising in response to plasma density and temperature gradients. This bootstrap current can be eliminated, though, by making the  stellarator ``quasi-isodynamic'' (QI). A QI field is, by definition, omnigenous and has poloidally closed contours of constant field strength. Trapped particles thus precess in the poloidal direction. In a QI field, the bootstrap current vanishes in the limit of low collision frequency \cite{helander2009,helander-2011}, which is helpful not only for avoiding current-driven instabilities, but also for facilitating island-divertor operation. A mathematical description of QI fields is given in \cref{subsec:QIconds}.

For these reasons, QI stellarators are a uniquely attractive option for a fusion reactor design, and are partially why the largest stellarator ever constructed, Wendelstein-7X in Germany, was designed as an attempt at a QI stellarator.

\subsection{Omnigenity and the second adiabatic invariant}

Mathematically, omnigenity can be understood through the so-called ``second adiabatic invariant'' ($\mathcal{J}$), which is a constant of motion for magnetically trapped particles. A particle in a toroidal magnetic field is defined as trapped if it is unable to enter regions with magnetic field strength $B\geq B_*$ within that field, and therefore will ``bounce'' along some fieldline $\alpha$ between points $l_1$ and $l_2$ at which the magnetic field strength $B=B_*$, and between which $B<B_*$. This particle's $\mathcal{J}$ can be written as
\begin{equation}
    \mathcal{J}=\int_{l_1}^{l_2} mv_\parallel \,\textrm{d}l,
    \label{eq:J}
\end{equation}
\noindent where $m$ is the particle's mass, $v_\parallel$ its velocity parallel to the magnetic field line, and $l$ is the geometric length along said field line. A trapped particle will in general be well confined if its second adiabatic invariant $\mathcal{J}$ is conserved when the particle remains on the same flux surface $s=\psi/\psi_\textrm{edge}$, where $\psi_\textrm{edge}$ is the toroidal magnetic flux, $\psi$, at the plasma boundary \cite{helander2014}. In the absence of an electric field within the flux surface, a particle with total energy $\mathcal{H} = m v^2/2$ has $v_\parallel$ given by
\begin{equation}
    v_\parallel = \pm \sqrt{2\frac{\mathcal{H}-\mu B}{m}},
    \label{eq:vpar}
\end{equation}
\noindent where $\mu = mv_\perp^2/2B$ is another invariant of the particle's motion, and $v_\perp$ is the particle's velocity perpendicular to the magnetic field line. Since $\cal H$, $\mu$, and $m$ are constants for collisionless particles, and because $v_\parallel=0$ when $B=B_*$, \cref{eq:vpar} can be substituted into \cref{eq:J} to yield
\begin{align}
    \mathcal{J} = \sqrt{\frac{2\mathcal{H}}{m}} \int_{l_1}^{l_2} \sqrt{1-\lambda B}\,\textrm{d}l
    \label{eq:J2}
\end{align}
\noindent where $\lambda\equiv 1/B_*=\mathcal{H}/\mu$. A particle's drift across field lines $\Delta\alpha$, and across flux surfaces $\Delta\psi \propto \Delta s$, are given by $\mathcal{J}$ as follows \cite{helander2014}:
\begin{equation}
    \Delta\alpha = -\frac{1}{Ze} \partial_\psi\mathcal{J}, \,\,\,\,\, \Delta\psi = \frac{1}{Ze} \partial_\alpha \mathcal{J},
    \label{eq:particle_drifts}
\end{equation}
\noindent where $Ze$ is the charge of the particle in question and $e$ is the elementary charge. We use the notation $\partial_x \equiv \partial/\partial x$ for any variable $x$.

Because the prefactor in \cref{eq:J2} is a constant of a particle's motion, it is often useful to think in terms of $\tilde{\mathcal{J}}(s,\alpha,\lambda) \equiv \mathcal{J} / \sqrt{2\mathcal{H}/m}$, which can now be understood as a property of the magnetic field, rather than a property of a particle. Note that while $\tilde{J}$ is still a function of $\lambda$ (a particle property), one can also express $\lambda=1/B_*$ (a magnetic field property). Evaluating $\tilde{\mathcal{J}}$ over various $s$ and $\lambda$ can thus be used to evaluate how well a field confines trapped particles. We can now connect $\mathcal{J}$ to the concept of omnigenity: in a perfectly omnigenous field, $\partial\tilde{\mathcal{J}}/\partial\alpha|_{s,\lambda}=0\,\,\forall \,\,s,\lambda,\alpha$.

\subsection[Maximum-J Configurations]{Maximum-\texorpdfstring{$\mathcal{J}$}{J} configurations}

Fields with the so-called ``maximum-$\mathcal{J}$'' property, where $\partial_\psi \mathcal{J} < 0$ (such that $\mathcal{J}$ decreases as a function of $\psi$) have supressed trapped electron mode (TEM) driven turbulence. This can be understood from the argument from \cite{Helander2012,proll2022}, which consider a plasma instability which moves a particle radially by $\Delta\psi>0$. The energy required to do so, $\Delta E$, can be found through the conservation of both $\mathcal{J}$ and the cumulative energy of the particle and instability ($\Delta\mathcal{H} + \Delta E = 0$):
\begin{equation}
    \Delta\mathcal{J} = \frac{\partial\mathcal{J}}{\partial\psi}\Delta\psi + \frac{\partial\mathcal{J}}{\partial\mathcal{H}}\Delta E = 0 \rightarrow \Delta E = -\frac{\partial\mathcal{J}/\partial\psi}{\partial\mathcal{J}/\partial\mathcal{H}}\Delta\psi.
    \label{eq:Jconserve}
\end{equation}
Because $\partial_\mathcal{H}\mathcal{J} < 0$ (see \cref{eq:J2}), we see that $\Delta E$ will be negative when $\partial_\psi\mathcal{J} < 0$. A negative $\Delta E$ means that the instability that caused the particle's radial displacement loses energy, and is therefore stabilized. For this reason, maximum-$\mathcal{J}$ is a desirable property in stellarators. Because $\mathcal{J}$ tends to be larger when $B$ is smaller (see \cref{eq:J2}), maximum-$\mathcal{J}$ is often also called minimum-$B$.

\subsection{Optimization}
Stellarator designs can be extremely complicated, and thus typically designed using numerical optimization algorithms, which require the use of a ``target function''. This target function should take some input (in this case, a stellarator design) and output a number that describes how ``good'' this stellarator design is. A novel choice of target function, which is described in \cref{subsec:QItarget}, was crucial to achieving the results presented.

There are many possible approaches to stellarator optimization \cite{henneberg2021}. For these optimizations, we employed the ideal magnetohydrodynamics (MHD) code \texttt{VMEC} \cite{VMEC}. In MHD, the magnetic field within a plasma is completely described by the shape of the plasma's boundary, as well as its current and pressure profiles. \texttt{VMEC} is thus able to take a plasma boundary as an input and output the corresponding magnetic field. This boundary shape is described by Fourier coefficients, which map toroidal and poloidal angles to physical coordinates $R$ and $Z$, taking the ``centre'' of the torus's ``doughnut-hole'' as the origin. 

These Fourier coefficients, $\textbf{x}$, are the inputs to our target function, $f$. We employed a nonlinear optimization algorithm, which uses numerically calculated finite-difference gradients to attempt to find the values of $\textbf{x}$ that result in the smallest possible target function output. 

\subsection{QI conditions}\label{subsec:QIconds}
\begin{figure}[ht]
    \begin{center}
        \includegraphics[width=0.55\linewidth]{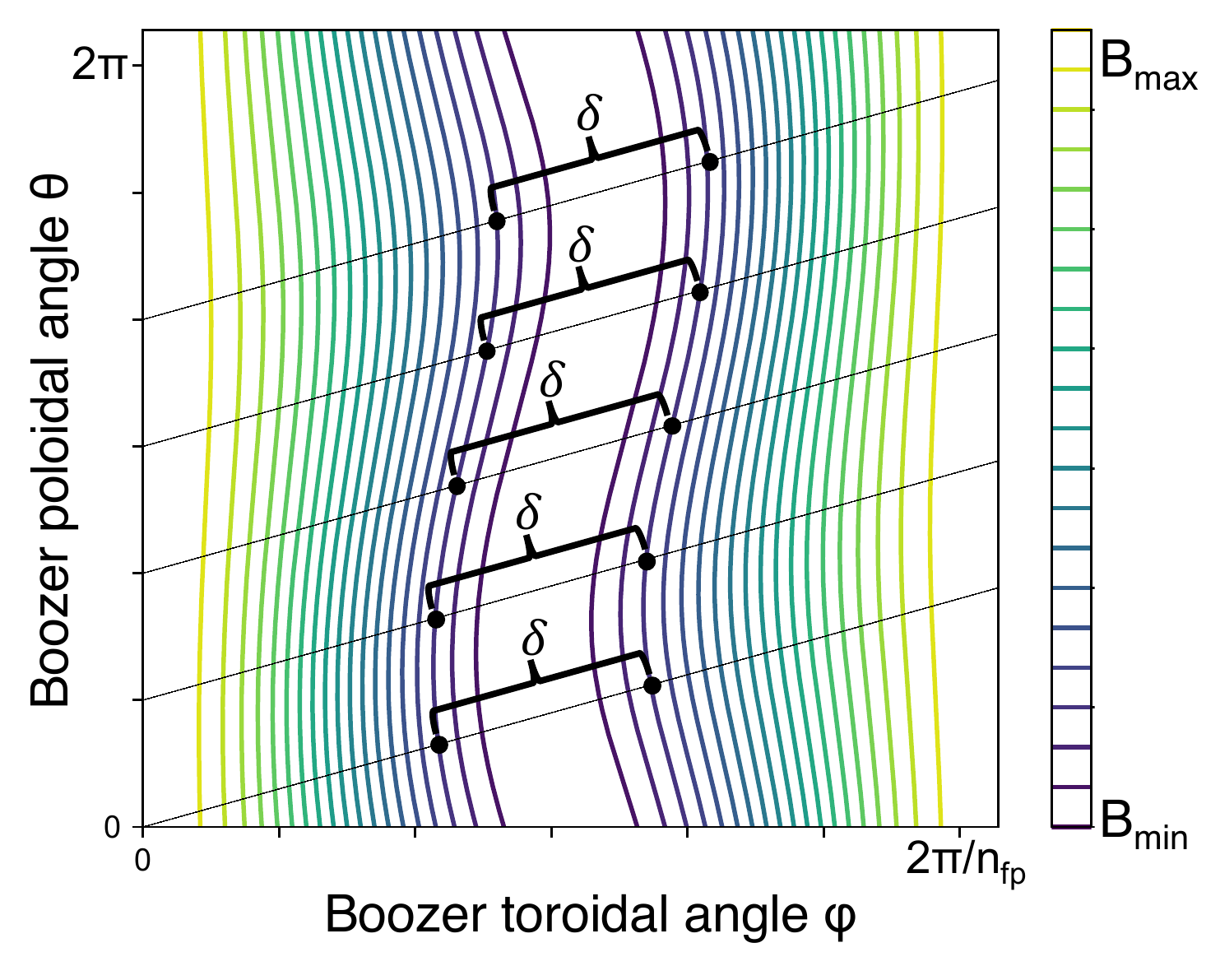}
        \captionsetup{justification=centering}
        \caption{A schematic of what contours of constant magnetic field strength (in Boozer coordinates) look like on a perfectly QI flux surface, constructed using the mapping from \cite{Landreman2012}. Fieldlines are shown as straight black lines.} 
        \label{fig:constructedQI}
    \end{center}
\end{figure}

In order to design a QI target function, one must understand the properties of a perfectly QI magnetic field. For the purposes of these optimizations, there are three conditions that describe the ``ideal'' perfectly QI magnetic field, corresponding to a geometry with $n_{fp}$ field-periods with $n_{fp}$ a natural number \cite{Cary:1997zzb}:
\begin{enumerate}[(1)]
    \item There should be only one minimum in magnetic field strength along any field line spanning a single field period. While this is not strictly necessary, it is a useful constraint to impose on a target function. \label{qic3}
    \item All contours of constant $B$ must close poloidally, meaning that
        \begin{enumerate}[(2i)]
            \item $\mathcal{B}_\textrm{min}(s)$ --- the minimum $B$ on each flux surface $s$ --- must be present exactly once on each field line along a single toroidal field-period, and\label{qic2i}
            \item $\mathcal{B}_\textrm{max}(s)$ --- the maximum $B$ on each flux surface $s$ --- must on every field line be located at the toroidal angle $\varphi=\{0,2\pi/n_{fp}\}$ in  Boozer coordinates \cite{Boozer1981}. In other words, contours of $\mathcal{B}_\textrm{max}(s)$ must be straight, vertical lines in the $\theta-\varphi$ plane.\label{qic2ii}
        \end{enumerate}\label{qic2}
    \item The ``bounce distance,'' $\delta$ in Boozer coordinates ($\theta,\varphi$) along a field line between consecutive points with equal $B=B_*$ (``branches'' \cite{helander2009}) should be independent of the field line label, $\alpha$, for any given flux surface $s$ ($\partial\delta/\partial\alpha|_{B,s}=0 \,\, \forall \,\, \alpha,B,s$).\label{qic1}
\end{enumerate}
A schematic of a QI field is shown in \cref{fig:constructedQI}.

A frequently studied subclass of omnigenity, known as quasisymmetry, requires that these contours be straight lines in Boozer coordinates. If $B$-contours are straight lines, and close poloidally, then the field has ``quasi-poloidal (QP) symmetry''. QP symmetry is impossible close to the magnetic axis \cite{helander2014}, and fields with precise QP symmetry have proven difficult to find through optimization \cite{landreman2022}, but because the conditions for QI are less strict, it offers a promising alternative.

%===========================================================================
\section{Target Function}
\label{sec:target}
%===========================================================================
In this work, we used the \texttt{SIMSOPT} optimization suite \cite{simsopt1,simsopt2}, to interface between \texttt{VMEC}, our new target function, and the trust-region reflective optimization algorithm in \texttt{scipy} \cite{scipy}. It is with these tools that we are able to optimize stellarator equilibria.

\subsection{QI target}\label{subsec:QItarget}
Using the conditions detailed in \cref{subsec:QIconds}, this target seeks to construct a closely related, perfectly QI field from an input field, and then penalizes the difference between the two (see \cref{eq:fQI}). At each optimization step, a new perfectly QI field is constructed.

\begin{figure*}
        \centering
        \begin{subfigure}[b]{0.475\textwidth}
            \centering
            \includegraphics[width=\textwidth]{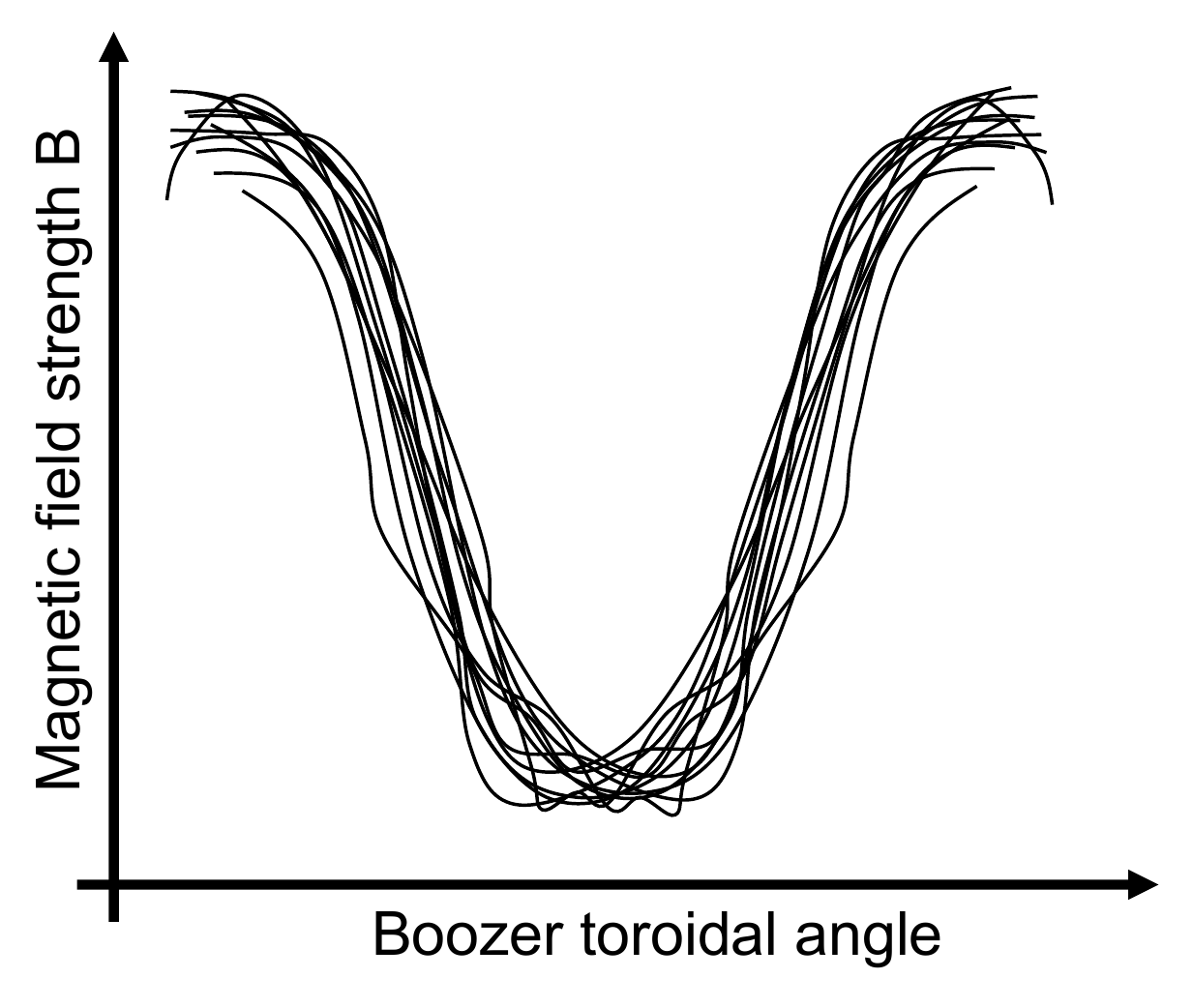}
            \caption[]%
            {{\small Actual $B$ along various field lines}}    
            \label{fig:bflex_orig}
        \end{subfigure}
        \hfill
        \begin{subfigure}[b]{0.475\textwidth}  
            \centering 
            \includegraphics[width=\textwidth]{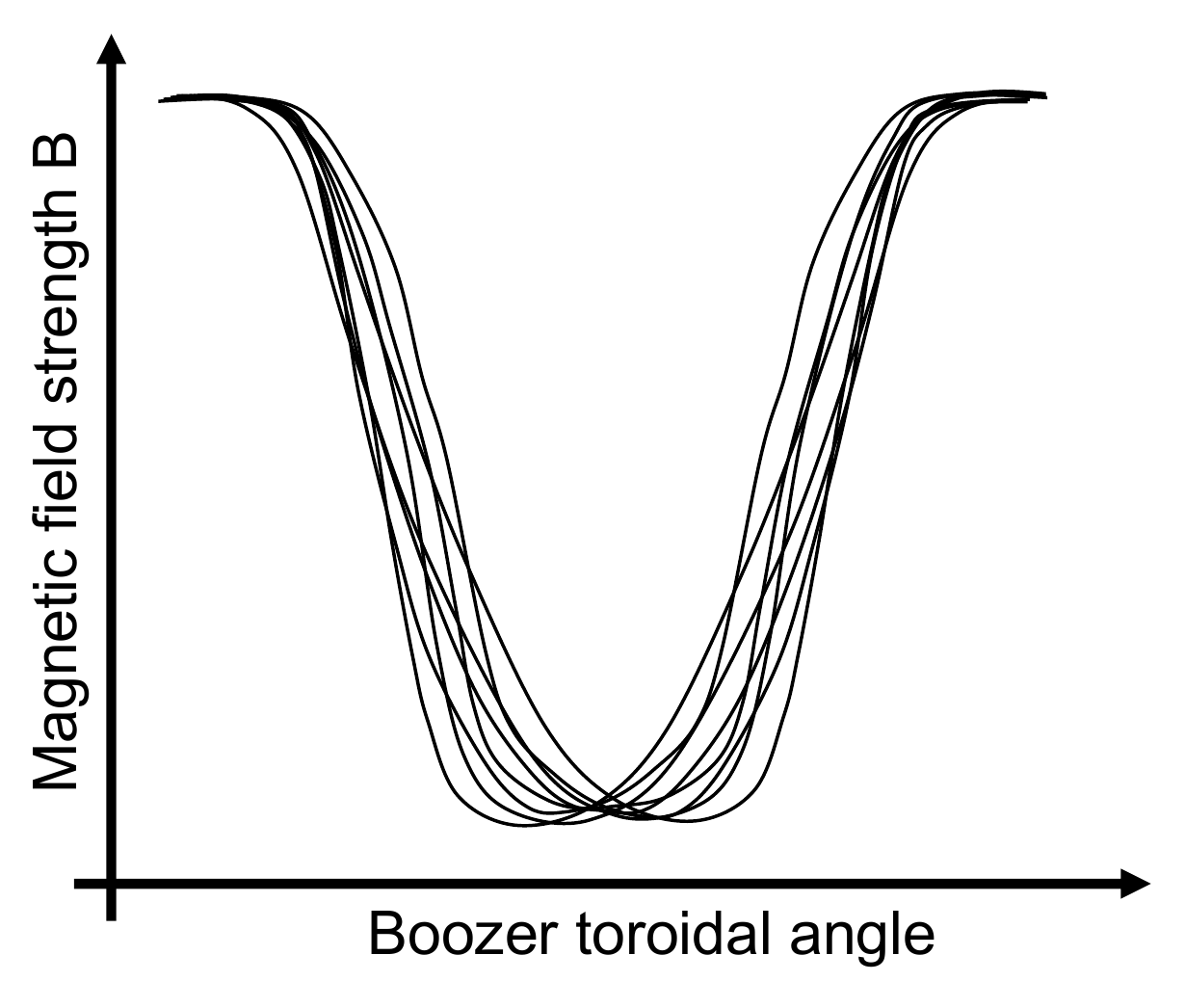}
            \caption[]%
            {{\small Modified QI $B$ along various field lines}}    
            \label{fig:bflex_omn}
        \end{subfigure}
        \captionsetup{justification=centering}
        \caption[]
        {\small An example of a set of non-QI wells (left), and a schematic of what a set of QI wells might look like (right).} 
        \label{fig:bflexs}
    \end{figure*}

In this work, we sample $n_\alpha$ field lines uniformly between $0$ and $2\pi$, and $n_\varphi$ values of $\varphi$ between $0$ and $2\pi/n_{fp}$ in Boozer space. We will call each $B(\varphi)|_{\alpha,s}$ a ``well.'' We then apply a set of three numerical transformations to these wells to create an artificial set of closely-related, perfectly QI wells. An example of what it might look like to turn a set of non-QI wells into a set of QI wells is shown in \cref{fig:bflexs}.
Note that there is no guarantee that the constructed QI wells (\cref{fig:bflex_omn}) are physically realisable. In fact, \cite{Cary:1997zzb} points out that an analytic QI field is impossible to achieve perfectly (although it may be possible to approach it arbitrarily closely). In this case, it does not matter that these wells are non-physical; what matters is that the difference between the original wells (\cref{fig:bflex_orig}) and the constructed wells (\cref{fig:bflex_omn}) describes how far the wells deviate from a QI field. Furthermore, as the input field becomes more QI, the target field will become closer to a ``physical'' field. Below are the steps taken to find a closely-related set of QI wells, analogous to those shown in \cref{fig:bflexs}.
\subsubsection*{Part 1: The Squash}
To construct a QI well from an input well, we first find the toroidal angle of its minimum magnetic field strength, $B_\textrm{min}$, and call it $\varphi_\textrm{min}$. Note that, in this input well, $B_\textrm{min}$ is not necessarily the same as $\mathcal{B}_\textrm{min}$ (the global minimum $B$ on the surface). In a perfectly QI field, Condition \ref{qic3} implies that $B_l = B(\varphi<\varphi_\textrm{min})$ and $B_r = B(\varphi>\varphi_\textrm{min})$ should be monotonically decreasing towards $\varphi_\textrm{min}$.

Because each well is defined by an array of points, we can enforce this condition by ``flattening'' any points that are non-monotonic. Specifically, if \texttt{B}$_\mathtt{l}$\texttt{[i]} is the array of field strengths for $\varphi<\varphi_\textrm{min}$ where $\texttt{i}=0$ corresponds to $\varphi=0$ and $\texttt{i}=n_i$ corresponds to $\varphi=\varphi_\textrm{min}$, we loop through \texttt{B}$_\mathtt{l}$\texttt{[i]} from \texttt{i}=$\{0,1,2,3,...,n_i-1\}$ and, if \texttt{B}$_\mathtt{l}$\texttt{[i+1]} \texttt{>} \texttt{B}$_\mathtt{l}$\texttt{[i]}, we set \texttt{B}$_\mathtt{l}$\texttt{[i+1]} \texttt{=} \texttt{B}$_\mathtt{l}$\texttt{[i]}. A similar operation for $B_r$ yields a well that is perfectly monotonic on either side of its minimum. An example of this operation is shown in \cref{fig:squashed}.

It is worth noting that, while this transformation does create monotonic wells, it is a very crude way of doing so. One can think of any number of ways to do this (for instance, one could sort the points in $B_l$ and $B_r$ in descending and ascending order respectively). It is unclear what the ``best'' way of doing this would be.

\subsubsection*{Part 2: The Stretch}

Our new well looks slightly more QI than the original, but further tranformations are still needed. Condition \ref{qic2} requires that $B_\textrm{min}=\mathcal{B}_\textrm{min}$, and $B(\varphi=0,2\pi/n_{fp})=\mathcal{B}_\textrm{max}$. To enforce this condition in our new well, we ``stretch'' $B_l$ and $B_r$ as follows:
\begin{equation}
\begin{aligned}
    B_l(\varphi) &\rightarrow \mathcal{B}_\textrm{min} + \frac{\mathcal{B}_\textrm{max} - \mathcal{B}_\textrm{min}}{B_l(\varphi=0) - B_\textrm{min}}(B_l(\varphi) - B_\textrm{min}), \\
    B_r(\varphi) &\rightarrow \mathcal{B}_\textrm{min} + \frac{\mathcal{B}_\textrm{max} - \mathcal{B}_\textrm{min}}{B_r(\varphi=2\pi/n_{fp}) - B_\textrm{min}}(B_r(\varphi) - B_\textrm{min}).
    \label{eq:stretch}
\end{aligned}
\end{equation}
We here define $\mathcal{B}_\textrm{min}$ and $\mathcal{B}_\textrm{max}$ to be the smallest and largest $B$ that are found on the flux surface.

After applying this transformation, these wells will satisfy Condition \ref{qic2}, as can be seen in \cref{fig:stretched}.

While we apply a ``linear'' stretch in this work, one could choose a different transformation that yields $\textrm{d}B/\textrm{d}\varphi=0$ at $\varphi=0,2\pi/n_{fp}$, possibly resulting in more realistic wells. It is unclear if this approach would work better.

The entirety of the $n_{fp}=1$ configuration's optimization (\cref{subsec:nfp1}), and the early stages of the $n_{fp}=2$ configuration's optimization (\cref{subsec:nfp2}) used the approximation that Boozer $\varphi=0$ was the same as \texttt{VMEC} $\phi=0$ so as to avoid a full Boozer coordinate transformation, although this approximation failed for the $n_{fp}=3$ optimization (\cref{subsec:nfp3}).

\subsubsection*{Part 3: The Shuffle}
Finally, we are ready to address Condition \ref{qic1}, which requires that all distances between consecutive branches are equal at constant $B_*$ for all $\alpha$. To do this, we first find the weighted-mean bounce distance for each field line on the surface in our new well, $\langle \delta \rangle_\alpha(B_i)$ for various values of $B_i$, which can be done numerically. The weightings $w$ in this mean, which were used to improve the smoothness of the optimization algorithm, are inversely proportional to the severity of the prior Squash and Stretch steps:

\begin{equation}
    w(\alpha) \propto \frac{1}{\int_0^{2\pi/n_{fp}} d\varphi \, \left(B_\textrm{(a)}(\alpha,\varphi) - B_\textrm{(c)}(\alpha,\varphi)\right)^2}
\end{equation}

\noindent where $B_\textrm{(a)}$ and $B_\textrm{(c)}$ are the Original and Stretched wells respectively. While at first glance these weights may appear to (possibly) diverge to infinity, which would be a problem. This is extremely unlikely, though, as it would require the maxima and minimum $B$ of the well to exactly equal $\mathcal{B}_\textrm{max}$ and $\mathcal{B}_\textrm{min}$. However, if this is a concern, would could simply modify $w(\alpha)$ to add some small number to the denominator.

With these weights, we can now write

\begin{equation}
    \langle \delta \rangle_\alpha = \sum_\alpha w(\alpha) (\varphi_2(B_i,\alpha) - \varphi_1(B_i,\alpha))
    \label{eq:dmean}
\end{equation}

\noindent for bounce points $\varphi_1(B_i,\alpha)$ and $\varphi_2(B_i,\alpha)$. To be QI, our well will satisfy $\delta(B_i,\alpha)=\langle \delta \rangle_\alpha(B_i)$ for all $B_i$ and $\alpha$. The motivation for using a weighted mean is to attempt to choose $\langle \delta \rangle_\alpha$ such that it most closely resembles the bounce distances of the original wells. Wells that have been Stretched and Squashed significantly may no longer resemble their original forms, and so their new bounce distances should not be as influential than bounce distances from wells that did not require significant Stretching or Squashing.

To force our new well into being perfectly QI, we simply need to transform each pair of bounce points as follows:
\begin{equation}
\begin{gathered}
    \varphi_1(B_i,\alpha) \rightarrow \varphi_1(B_i,\alpha) + \Delta\varphi_1, \\
    \varphi_2(B_i,\alpha) \rightarrow \varphi_2(B_i,\alpha) + \Delta\varphi_2.
\end{gathered}
\end{equation}
There are many methods by which $\Delta\varphi_1$ and $\Delta\varphi_2$ can be chosen, so long as 
\begin{align}
    \delta(B_i,\alpha) - \Delta\varphi_1 + \Delta\varphi_2 = \langle \delta \rangle_\alpha(B_i),
\end{align}
\noindent but they should ideally be chosen so as to minimize the difference between the shuffled well and the stretched well. One must also take care that stellarator symmetry is preserved (if the optimizations are stellarator symmetric), and also must ensure
\begin{eqnarray}
    & \varphi_1(B_i) < \varphi_2(B_i),  \label{eq:wc1} \\
    & \varphi_1(B_i) < \varphi_1(B_j),  \label{eq:wc2} \\ 
    & \varphi_2(B_i) > \varphi_2(B_j),  \label{eq:wc3} \\
    & B(\varphi=0)=B(\varphi=2\pi/n_{fp})=\mathcal{B}_\textrm{max} \label{eq:wc4}
\end{eqnarray}

\noindent for $B_j<B_i$. In this work, we shuffled the points around $B_\textrm{min}$ first, and moved ``up'' the well until $B_\textrm{max}$ had also been shuffled. For each set of bounce points, we chose $\Delta\varphi_1 = \Delta\varphi_2 = (\langle\delta\rangle_\alpha - \delta)/2$. If this transformation violated Eqs. (\ref{eq:wc2}) or (\ref{eq:wc3}), both points were shifted by $\varphi_1(B_i) - \varphi_1(B_j) + 10^{-5}$ or $\varphi_2(B_j) - \varphi_2(B_i) - 10^{-5}$ respectively, so as to correct this violation.

\begin{figure*}[ht]
        \centering
        \begin{subfigure}[b]{0.475\textwidth}
            \centering
            \includegraphics[width=\textwidth]{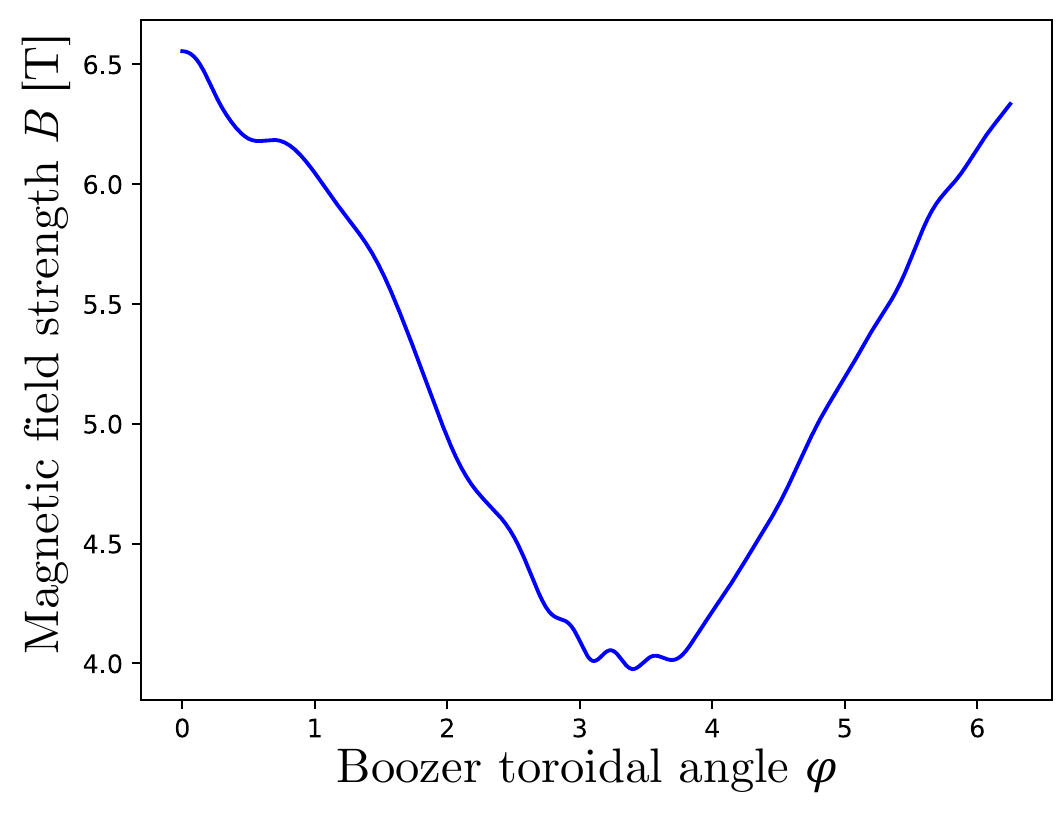}
            \caption[]%
            {{\small Original}}    
            \label{fig:original}
        \end{subfigure}
        \hfill
        \begin{subfigure}[b]{0.475\textwidth}  
            \centering 
            \includegraphics[width=\textwidth]{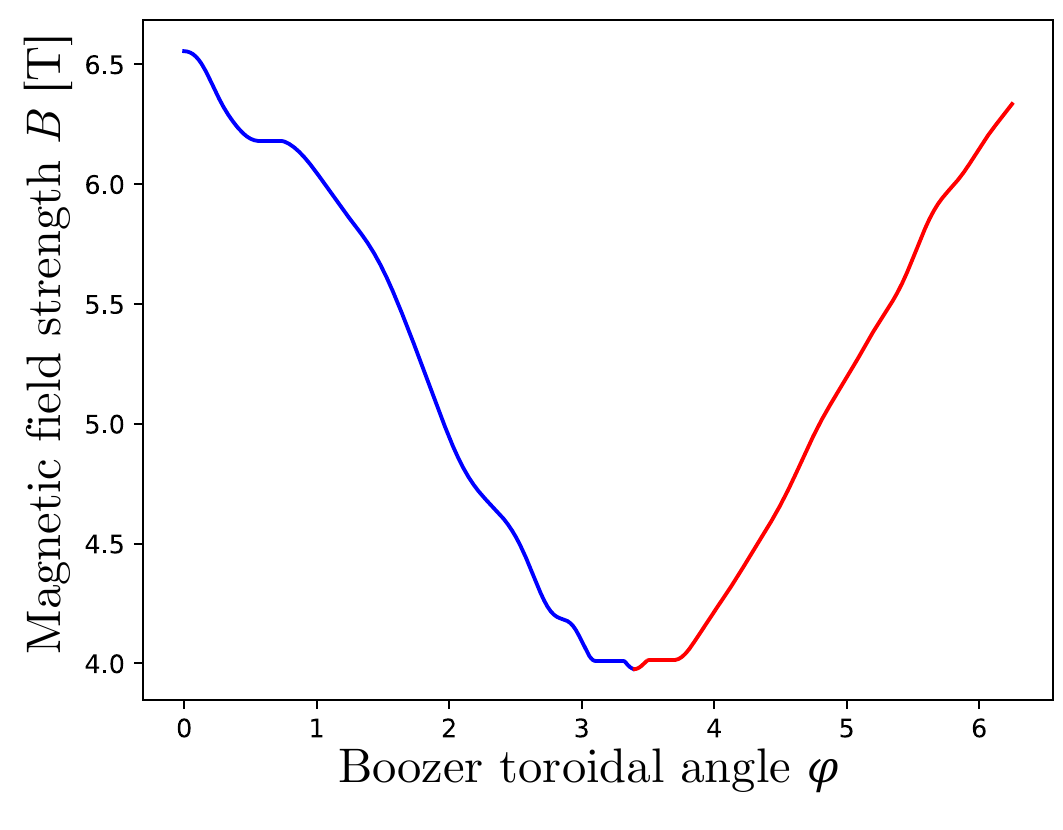}
            \caption[]%
            {{\small Squashed}}    
            \label{fig:squashed}
        \end{subfigure}
        \vskip\baselineskip
        \begin{subfigure}[b]{0.475\textwidth}   
            \centering 
            \includegraphics[width=\textwidth]{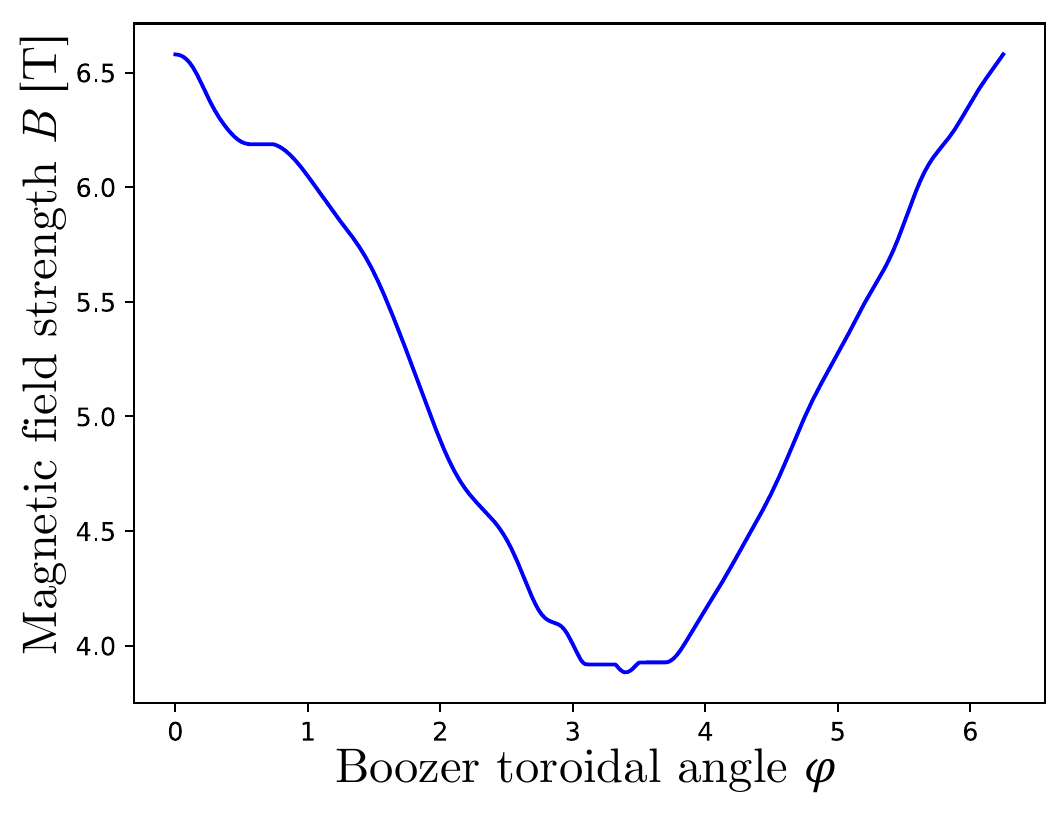}
            \caption[]%
            {{\small Stretched}}   
            \label{fig:stretched}
        \end{subfigure}
        \hfill
        \begin{subfigure}[b]{0.475\textwidth}   
            \centering 
            \includegraphics[width=\textwidth]{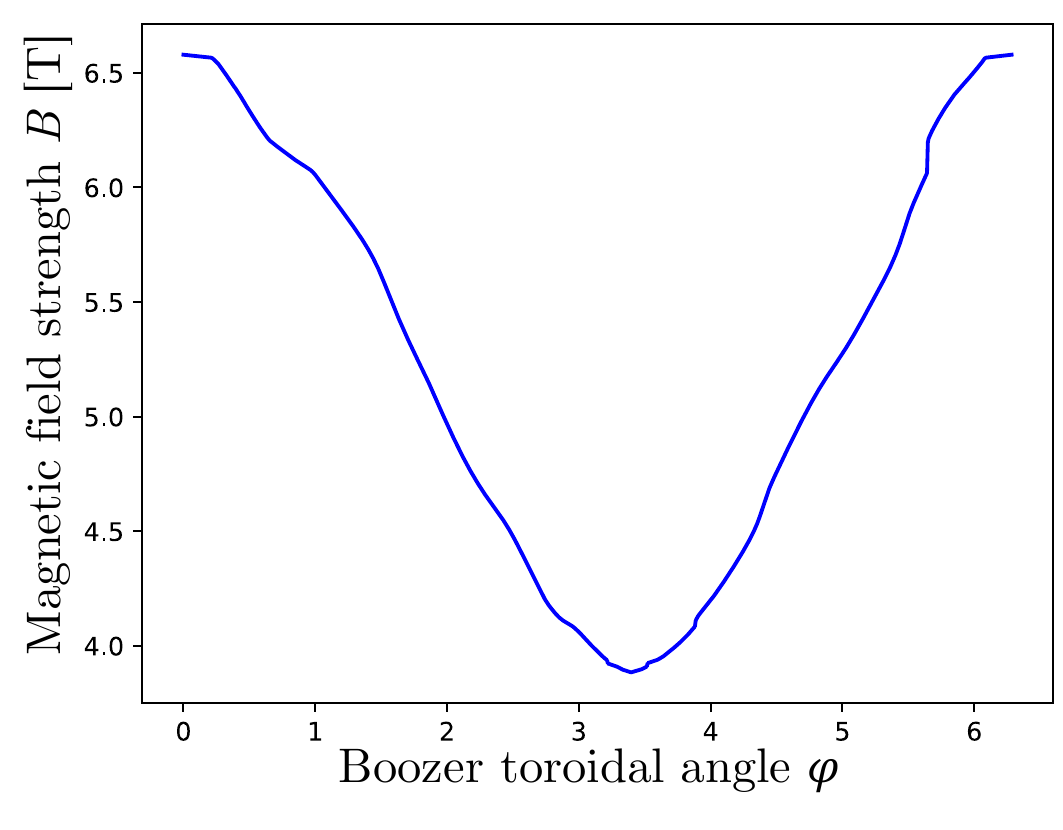}
            \caption[]%
            {{\small Shuffled}} 
            \label{fig:shuffled}
        \end{subfigure}
        \captionsetup{justification=centering}
        \caption[]
    {\small The various transformations to get from $B$ to $B_\textrm{QI}$. Note that $B_l$ and $B_r$ in (b) are coloured blue and red respectively.}
    \label{fig:transform}
\end{figure*}

After this modification, the resulting field, which we call $B_\textrm{QI}(s,\alpha,\varphi)$ should be completely QI. An example of such a well is shown in \cref{fig:shuffled}.
We have now constructed a ``target field'' that is close to our input field. To penalize deviations from QI, we finally construct our penalty function, $f$, as the difference between the scaled original field and the scaled constructed, QI field:
\begin{align}
    f_\textrm{QI}(s) = \frac{n_{fp}}{4\pi^2} \frac{1}{(\mathcal{B}_\textrm{max} - \mathcal{B}_\textrm{min})^2} \int_0^{2\pi} \textrm{d}\alpha \int_0^{2\pi/n_{fp}}\textrm{d}\varphi \, \left(\tilde{B}(s,\alpha,\varphi) - \tilde{B}_\textrm{QI}(s,\alpha,\varphi)\right)^2,
    \label{eq:fQI}
\end{align}
\noindent where the prefactor $1/(\mathcal{B}_\textrm{max} - \mathcal{B}_\textrm{min})^2$ makes the penalty dimensionless.

\subsection{Elongation target}
Recent efforts to find QI fields have resulted in very elongated flux surfaces, such that the semi-minor radius ($a$) of a poloidal cross-section is significantly smaller than its semi-major radius ($b$) \cite{jorge2022,mata2022}. In fact, close to the magnetic axis QI can be achieved to arbitrary accuracy by making the flux surfaces infinitely elongated in the direction perpendicular to the curvature vector. The magnetic drift then becomes tangential to these surfaces almost everywhere. High elongation is however undesirable in stellarators (as we discuss in \cref{sec:discussion}), and so we included a penalty function to keep elongation under control.

Because stellarator boundaries are three-dimensional, the shape of a poloidal cross-section depends on the toroidal angle $\phi$. Hence, to calculate elongation, we must sample several cylindrical angles, and expect to find a different elongation at each angle.

Crucially, the angle (relative to the magnetic axis) at which the poloidal cross-section is taken is also an important factor when calculating elongation. To understand this, we first define the vector normal to the poloidal cross-section as $\mathbf{n_x}$, and the vector tangent to the magnetic axis as $\mathbf{t_a}$. If $\mathbf{n_x}$ and $\mathbf{t_a}$ are not parallel, then the shape of the cross-sections will change. Because \texttt{VMEC} uses the cylindrical angle $\phi$ to define its boundary, $\mathbf{n_x}$ and $\mathbf{t_a}$ will necessarily not be parallel in configurations with a non-planar magnetic axes. We hence must take care that, for a given cross-section which intersects the magnetic axis at a point in real space $\mathbf{p_a}$ at \texttt{VMEC} angle $\phi_a$, each point that defines the edge of said cross-section $\mathbf{p_x}(\phi_x)$ satisfies the relation
\begin{equation}
    \mathbf{t_a}(\phi_a) \cdot [\mathbf{p_a}(\phi_a) - \mathbf{p_x}(\phi_x)] = 0.
    \label{eq:elongcondition}
\end{equation}
We define our cross-sections by numerically finding points on the boundary at various values of $\phi_x$ using \cref{eq:elongcondition} for several \texttt{VMEC} $\theta$ coordinates between $0$ and $2\pi$.

Because these poloidal cross-sections are not, in general, ellipses, and can take any number of unusual shapes, the definition of $a$ and $b$ are somewhat ambiguous. We address this problem by defining ``effective'' semi-major and -minor axes by finding the cross-section's circumference, $C_\sigma$, and its area, $A_\sigma$, and using them find the ``effective elongation'' $\varepsilon \equiv b/a$ of the ellipse with the same circumference and area. In this work, we penalized only the maximum elongation.

\subsection{Additional targets}
Two other properties that the optimizer could (and did) exploit to make $f_\textrm{QI}\rightarrow0$ are arbitrarily large aspect ratios $A$, and/or mirror ratios 
\begin{equation}
\Delta=\frac{\mathcal{B}_\textrm{max} - \mathcal{B}_\textrm{min}}{\mathcal{B}_\textrm{max} + \mathcal{B}_\textrm{min}}.
\label{eq:mirr}
\end{equation}
These are undesirable properties in a stellarator, so we set our penalty function to be
\begin{align}
    f = w_\textrm{QI}\sum_s f_\textrm{QI}(s) + 
    w_\Delta \textrm{max}(0,\Delta-\Delta_*)^2 + 
    w_\varepsilon \textrm{max}(0,\varepsilon-\varepsilon_*)^2 +
    w_A \textrm{max}(0,A-A_*)^2,
    \label{eq:f}
\end{align}
\noindent where $\Delta_*$, $\varepsilon_*$, and $A_*$ are the maximum acceptable ``cut-in'' values of the mirror ratio, maximum elongation, and aspect ratio respectively, and are chosen somewhat arbitrarily. By setting $w_\textrm{QI} \ll w_\Delta \sim w_\varepsilon \sim w_A$, we ensure the results of these optimizations do not exceed these cutoffs.

The mirror ratio $\Delta$ is calculated by numerically finding $\mathcal{B}_\textrm{max}$ and $\mathcal{B}_\textrm{min}$ on flux surface $s=1/51$, as this was the lowest flux-surface calculated in VMEC in the early stages of these optimizations. The elongation $\varepsilon$ is calculated by finding the approximate boundary cross-sections on planes normal to the magnetic axis at various physical angles $\phi$, and finding the elongation of the ellipse with the same area-to-circumference ratio of each cross-section. \texttt{VMEC}'s aspect ratio is taken as $A$.

Note that during the Stretch step, \cref{eq:stretch} allows a built-in way to limit the mirror ratio, which must be done when optimizing for QI. On some surface reasonably near the axis, simply replacing $\mathcal{B}_\textrm{min}$ and $\mathcal{B}_\textrm{max}$ with values that result in a desired mirror ratio may suffice, although this was not done in this work.

\section{Initial Conditions}
%===========================================================================
The optimization method presented in this work is sensitive to the choice of initial conditions. Because of this, we chose initial conditions by exploiting the ``near-axis expansion'' \cite{landreman2019,jorge2020,plunk2021}, which allows us to construct a first-order plasma boundary given by a rotating ellipse, based around a carefully chosen magnetic axis.
According to the near-axis expansion, magnetic axes in QI fields should have zero curvature at points where field is at its minimum and maximum \cite{mata2022}. There is also reason to believe \cite{mata2022} that careful control over the axis' torsion can dramatically improve these constructions, although it is not accounted for in this approach.

To find the $n_{fp}=2$ configuration presented, we looked to section 6 in \cite{mata2022} as a starting point, which includes a two field-period configuration constructed from the near-axis expansion. Realizing this configuration with high fidelity requires a large number of poloidal (\texttt{mpol}) and toroidal (\texttt{ntor}) modes in \texttt{VMEC} (around 10 and 90 respectively), which is problematic for optimization. Hence, we simply truncated our Fourier spectrum to remove all Fourier coefficients with mode numbers > 2. The result was a configuration with very poor QI, but with a similar boundary and axis shape to its high-mode counterpart. The initial condition for the $n_{fp}=3$ configuration was a \texttt{VMEC} input file from the previously completed $n_{fp}=2$ optimization, with the number of field-periods increased, truncating the boundary's Fourier spectrum to include only the first \texttt{mpol}$=$\texttt{ntor}$=2$.

In the next section, we describe a novel approach was used to generate the initial condition for the $n_{fp}=1$ configuration presented. This initial condition was effective for $n_{fp}=1$, but not for $n_{fp}>1$.

\subsection{Heuristic construction}\label{subsec:heuristic}
In this section, we present a heuristic model for the boundary shape
of a configuration that is close to quasi-isodynamic or quasi-poloidal
symmetry. The goal is to find a boundary shape with very small number
of Fourier modes in the usual representation in cylindrical coordinates,
which could be used to initialize optimization. At the same time,
the model may give some insight as to the character of the space of
solutions, e.g. how they could be parameterized. It also indicates
the minimal set of Fourier modes to use for the first optimization
stage. The approach here is not rigorous, as the method in \cite{Plunk19} is, but has the advantages of having no
differential equations to solve and giving a boundary shape described
by very few Fourier modes. The principles of the model are as follows:
\begin{itemize}
\item The maximum $B$ will occur at $\phi=0$ and the minimum $B$ will
occur at $\phi=\pi/n_{fp}$.
\item The curvature of the magnetic axis will need to vanish at these values
of $\phi$, or else $\mathcal{B}_{\max}$ and $\mathcal{B}_{\min}$ on flux surfaces to
first order in $\sqrt{\psi}$ would be points instead of poloidally
closed curves. 
\item The flux surface shapes surrounding the axis will be ellipses that
rotate as $\phi$ increases. The minor axis of the ellipse will
approximately align with the axis curvature vector, to minimize the poloidal variation of $B$. This can be understood from the near-axis relation
$B_{1}=\kappa X_{1}B_{0}$, derived in \cite{GarrenBoozer1}, where $B = B_0 + r B_1 + O(r^2)$, $B_0(\varphi)$ is the field strength on axis, $r \propto \sqrt{\psi}$ is a surface label, $\kappa(\varphi)$ is the axis curvature, and $r X_1(\theta,\varphi)$ is the extent of the surfaces in the direction of the axis normal vector.
We want $B_{1}$ to be small so it does not strongly modify the quasi-poloidally
symmetric field associated with the mirror in $B_{0}$.
\item The magnetic axis will have some torsion, and this torsion and the
rotating elongation will contribute constructively to $\iota$. 
\item The cross-sectional area of the flux surfaces will vary toroidally,
in such a way as to give the desired mirror ratio. 
\end{itemize}
We proceed by first finding a magnetic axis with the desired
points of vanishing curvature, then defining a surface surrounding
it with the appropriate rotating elongation, and finally adjusting
this surface to provide a mirror term. 

First, for the magnetic axis shape, we adopt the minimal model
\begin{equation}
R_{ax}\left(\phi\right)=R_{0,0}+R_{0,2}\cos\left(2n_{fp}\phi\right),\;\;Z_{ax}\left(\phi\right)=Z_{0,2}\sin\left(2n_{fp}\phi\right)\label{eq:axis_shape}
\end{equation}
Note the factors of 2, introduced here because the symmetry of the
flux surface shapes will be different from the symmetry of the axis
itself. For instance a configuration with $n_{fp}=1$ has a single
$\mathcal{B}_{\max}$ and single $\mathcal{B}_{\min}$, meaning there are two points of
vanishing curvature on the axis, and so generally the axis can be
symmetric under rotation by $\pi$. Let $\mathbf{r}\left(\phi\right)=\left[R_{ax}\cos\phi,R_{ax}\sin\phi,Z_{ax}\right]$
denote the position vector along the axis. The axis curvature can
be computed from $\kappa=\left|\mathbf{r}'\times\mathbf{r}''\right|/\left|\mathbf{r}'\right|^{3}$,
where primes denote $d/d\phi$. For the curve described by \cref{eq:axis_shape},
the curvature at $\phi=0$ and $\phi=\pi/n_{fp}$ is
\begin{equation}
\kappa=\frac{\left|R_{0,0}+\left(1+4n_{fp}^{2}\right)R_{0,2}\right|}{\left(R_{0,0}+R_{0,2}\right)^{2}+4n_{fp}^{2}Z_{0,2}^{2}}.
\end{equation}
Therefore the curvature vanishes at these points when
\begin{equation}
R_{0,2}=-\frac{R_{0,0}}{1+4n_{fp}^{2}}.
\end{equation}
Note that there is no constraint on the coefficient $Z_{0,2}$, which can be
adjusted to vary the axis torsion and hence vary $\iota$. If desired,
the number of Fourier modes in the axis shape \cref{eq:axis_shape}
could be increased, and $\kappa\left(\phi\right)$ could be made to
vanish to higher order \citep{mata2022,jorge2022}, but the expressions
here are sufficient for the minimal model here.

Next, we establish a rotating ellipse surrounding the axis:
\begin{eqnarray}
R\left(\vartheta,\phi\right)
&=&
R_{ax}\left(\phi\right)+\hat{R}\left(\vartheta\right)\cos\left(kn_{fp}\phi\right)+\hat{Z}\left(\vartheta\right)\sin\left(kn_{fp}\phi\right),\label{eq:ellipse}
\\
Z\left(\vartheta,\phi\right)
&=&
Z_{ax}\left(\phi\right)-\hat{R}\left(\vartheta\right)\sin\left(kn_{fp}\phi\right)+\hat{Z}\left(\vartheta\right)\cos\left(kn_{fp}\phi\right),
\end{eqnarray}
where $\hat{R}=a\cos\vartheta$, $\hat{Z}\left(\vartheta\right)=b\sin\vartheta$,
$a$ is the semi-major radius of the ellipse (half the major diameter) and
$b$ is the semi-minor radius of the ellipse. 
The variable $\vartheta$ used is not a true poloidal angle, and can be understood as the angle about the rotating ellipse, in the ellipse's ``rest frame,'' such that $\vartheta=0$ and $\vartheta=\pi/2$ are always along its semi-major and -minor axes respectively.
The constant $k$ describes
the number of rotations of the ellipse per field-period; we will consider
$k=1$ for simplicity, but other values could be considered. Also
for simplicity we have assumed in this construction that the ellipse
rotates at a uniform speed with respect to $\phi$, and that the elongation
is independent of $\phi$. Neither of these choices is likely to be
precisely optimal, but they are good enough for this simplistic model. 

We now switch to a true poloidal angle $\theta=\vartheta-n_{fp}\phi$,
and introduce the elongation $\varepsilon=a/b$. If $\varepsilon>1$, then the ellipses
will be oriented so the thin dimension is roughly aligned with the curvature
vector, which will minimize $B_{1}$. Applying some trigonometric
identities, we arrive at
\begin{eqnarray}
R\left(\theta,\phi\right)
&=&
R_{ax}\left(\phi\right)+\frac{\left(\varepsilon-1\right)b}{2}\cos\left(\theta+2n_{fp}\phi\right)+\frac{\left(\varepsilon+1\right)b}{2}\cos\theta,
\\
Z\left(\theta,\phi\right)
&=&
Z_{ax}\left(\phi\right)-\frac{\left(\varepsilon-1\right)b}{2}\sin\left(\theta+2n_{fp}\phi\right)+\frac{\left(\varepsilon+1\right)b}{2}\sin\theta.
\end{eqnarray}

It is convenient to compute $a$ and $b$ in terms of an effective
aspect ratio $A$. The average minor radius conventionally used in
the stellarator community, $a_{eff},$ is defined such that $\pi a_{eff}^{2}$
equals the toroidal average of the cross-sectional area in the $R-Z$
plane. In our case, this area is $\pi ab$, so $a_{eff}=\sqrt{ab}$.
Then the aspect ratio $A$ is approximately $A=R_{0,0}/a_{eff}=R_{0,0}/\left(b\sqrt{\epsilon}\right)$,
so
\begin{equation}
b=\frac{R_{0,0}}{A\sqrt{\epsilon}},
\hspace{1in} 
a=\frac{R_{0,0}\sqrt{\epsilon}}{A}.
\label{eq:ab_from_aspect}
\end{equation}

The remaining task is to adjust the surface shape to create a mirror
term.This is achieved by adding a toroidal variation to the size of
the ellipse; flux conservation then implies that an increase in the
area of the ellipse corresponds to a proportional decrease in $B$.
Specifically, we add the term $-\xi a\cos\theta\cos(n_{fp}\phi)$ to
$R$, and add $-\xi b\sin\theta\cos(n_{fp}\phi)$ to $Z$, for a constant
$\xi$. The motivation for these terms is that they result in
the $\phi=0$ cross-sectional dimensions being reduced by a factor
$\left(1-\xi\right)$,
\begin{equation}
R\left(\theta,0\right)=R_{ax}\left(0\right)+a\left(1-\xi\right)\cos\theta,\;\;Z\left(\theta,0\right)=b\left(1-\xi\right)\sin\theta,
\end{equation}
while the $\phi=\pi/n_{fp}$ cross-sectional dimensions are increased
by a factor $\left(1+\xi\right)$:
\begin{equation}
R\left(\theta,\pi/n_{fp}\right)=R_{ax}\left(\pi/n_{fp}\right)+a\left(1+\xi\right)\cos\theta,\;\;Z\left(\theta,\pi/n_{fp}\right)=b\left(1+\xi\right)\sin\theta.
\end{equation}
(The choice of terms to add to $R$ and $Z$ to achieve this kind
of effect is not unique, but the choice here is convenient as it introduces
no $\theta-3n_{fp}\phi$ Fourier modes.) Flux conservation gives
\begin{equation}
\mathcal{B}_{\max}ab\left(1-\xi\right)^{2}=\mathcal{B}_{\min}ab\left(1+\xi\right)^{2}.
\end{equation}
Rearranging, we can write the relative size of the mirror term (see \cref{eq:mirr}) as
\begin{equation}
\Delta=\frac{2\xi}{1+\xi^{2}},
\end{equation}
which can be inverted to give
\begin{equation}
\xi=\frac{1-\sqrt{1-\Delta^{2}}}{\Delta}.\label{eq:epsilon_from_Delta}
\end{equation}

Applying a trigonometric identity for the new terms associated with
the mirror field, we finally arrive at
\begin{eqnarray}
R\left(\theta,\phi\right) & = & R_{0,0}+R_{0,2}\cos\left(2n_{fp}\phi\right)+\frac{\left(\varepsilon-1\right)b}{2}\cos\left(\theta+2n_{fp}\phi\right)+\frac{\left(\varepsilon+1\right)b}{2}\cos\theta\nonumber \\
 &  & -\frac{\xi a}{2}\cos\left(\theta+n_{fp}\phi\right)-\frac{\xi a}{2}\cos\left(\theta-n_{fp}\phi\right),\label{eq:R_final}
\end{eqnarray}
\begin{eqnarray}
Z\left(\theta,\phi\right) & = & Z_{0,2}\sin\left(2n_{fp}\phi\right)-\frac{\left(\varepsilon-1\right)b}{2}\sin\left(\theta+2n_{fp}\phi\right)+\frac{\left(\varepsilon+1\right)b}{2}\sin\theta\nonumber \\
 &  & -\frac{\xi b}{2}\sin\left(\theta+n_{fp}\phi\right)-\frac{\xi b}{2}\sin\left(\theta-n_{fp}\phi\right).
\end{eqnarray}
This expression is supplemented with \cref{eq:ab_from_aspect,eq:epsilon_from_Delta} to compute $a$, $b$, and $\xi$
in terms of $A$, $\varepsilon$, $\Delta$, and $R_{0,0}$.

The basic input parameters to this boundary shape are the average
major radius $R_{0,0}$, the aspect ratio $A$, the elongation $\varepsilon$,
the mirror ratio $\Delta$, and the vertical extent of the axis $Z_{0,2}$.
Presumably, larger $\Delta$ helps with obtaining good quality of
QP symmetry, since for a larger mirror ratio, $B_{1}$ will distort
the vertical $B$ contours less. But if $\Delta$ is too large, the
reduced $B$ at $\phi=\pi/n_{fp}$ becomes low enough that the confinement
degrades there. Increasing $\varepsilon$ helps with reducing $B_{1}$, thereby
improving the QP symmetry. Also, increasing $\varepsilon$ and $Z_{0,2}$ will
increase $\iota$, which is good since it presumably increases the
$\beta$ limit, but at the expense of increased shaping which may
degrade the QP symmetry and make coils more difficult. Since both
$\varepsilon$ and $Z_{0,2}$ should strongly affect $\iota$, it appears that
there is significant flexibility in $\iota$. 

One property of this construction that is noteworthy is that $n=2 n_{fp}$
modes are required, both for $m=0$ and $m=1$. This means that if
one tries to initially optimize in a small parameter space, consisting
of very few Fourier modes, at least these modes must be included.
In particular one should not truncate the parameter space to only
modes with $n\le n_{fp}$.

A natural generalization of this model is the following. The surfaces
can be ellipses in the plane perpendicular to the magnetic axis, with
the minor axis of the ellipse oriented along the normal vector of
the magnetic axis. The boundary surface can then be computed in standard
cylindrical coordinates using the method of section 4.2 in \cite{landreman2019}. This variant of the procedure has the advantage of being
slightly more rigorous in the minimization of $B_{1}$ and in the
determination of the ellipse's rotation at each $\phi$, but at the
cost of more Fourier modes in the boundary shape.

%%%%%%%%%%%%%%%%%%%%%%%%%%%%%%%%%%%%%%%
%%%%%%%%%%%%%%%%%%%%%%%%%%%%%%%%%%%%%%%

%===========================================================================
\section{Results}

\begin{figure}[ht]
    \begin{center}
        \includegraphics[width=0.995\linewidth]{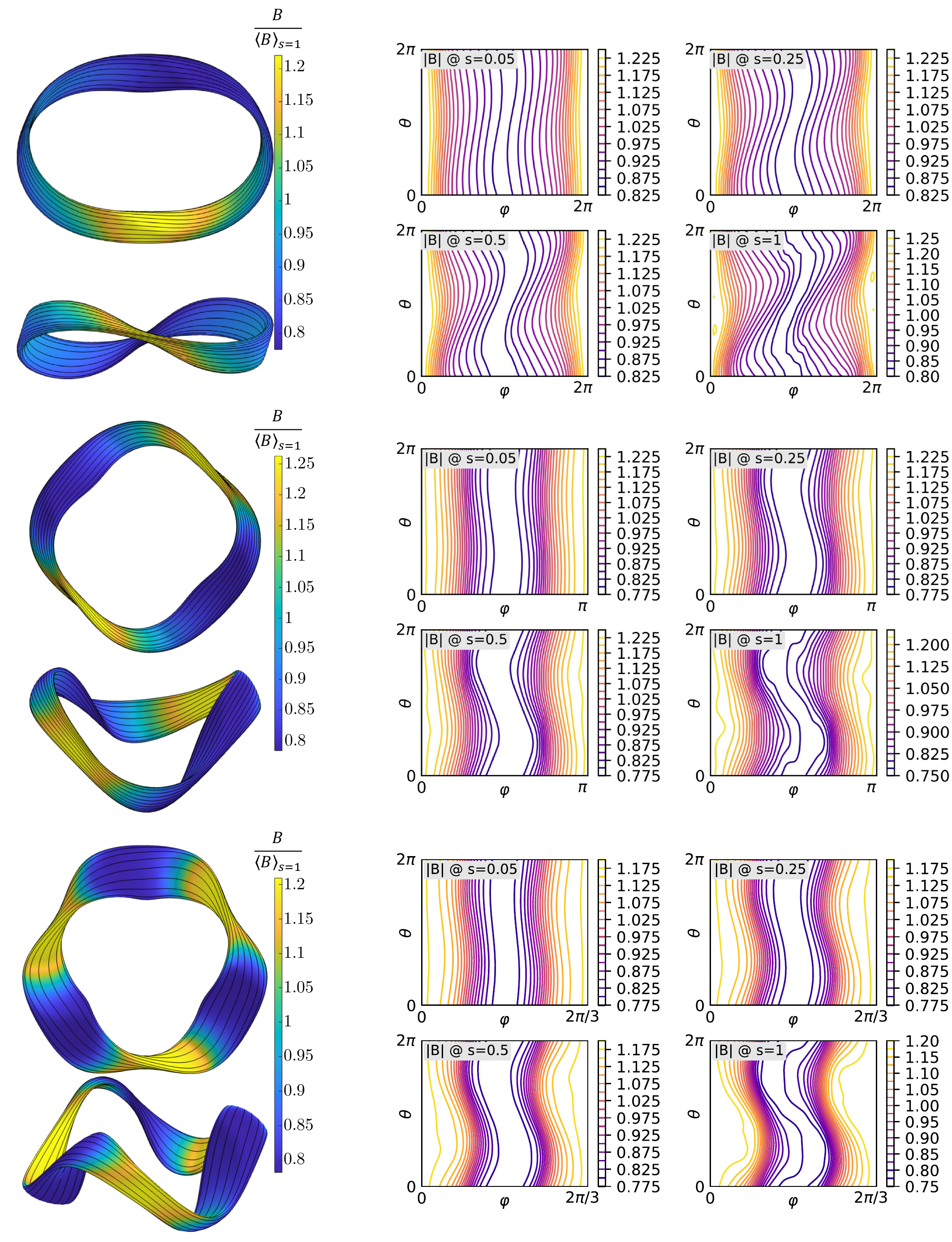}
        \captionsetup{justification=centering}
        \caption{
        New QI configurations with one (top), two (middle), and three (bottom) field-periods.
        \textbf{Left:} two views (top/bottom) of the boundary of QI configurations. The colour map represents a normalized magnetic field strength, with fieldlines shown in black.
        \textbf{Right}: magnetic field strength contours on four flux surfaces, in Boozer coordinates, for each corresponding configuration.} 
        \label{fig:3dplots}
    \end{center}
\end{figure}

To demonstrate the robustness of this approach, we present three different optimized vacuum configurations which show excellent QI properties. The three-dimensional representation of these configurations' boundaries and the corresponding magnetic field strength can be seen in \cref{fig:3dplots}. We evaluated all of these fields using the code \texttt{SPEC} \cite{spec1,spec2} and found that the \texttt{VMEC} solutions are undisturbed by magnetic islands. In all cases, the rotational transform $\iota$ is larger at the magnetic axis than at the plasma boundary.

When performing these optimizations, we began by optimizing the non-zero modes of the initial condition. We then gradually optimized larger \texttt{VMEC} modes, increasing the toroidal mode number \texttt{ntor} twice as quickly as the poloidal mode number \texttt{mpol}. This is motivated by the fact that QI equilibria generated from near-axis expansion solutions generally require more toroidal modes than poloidal modes.

All configurations presented in this paper, the QI target function presented above, and most of the codes used to create this paper's figures can be found at \cite{qi_archive}. Some plotting routines included were taken from \cite{pqs_archive}.
\label{sec:results}
%===========================================================================
\subsection{One field-period}\label{subsec:nfp1}
We first present a precisely QI configuration with a single field-period, inspired by the success of the configuration published in \cite{jorge2022}. At first, this configuration was optimized with no rotational transform target, but the resulting field was found to have magnetic islands when run through \texttt{SPEC}, due to its rotational transform passing through the low-order rational 3/5. Further optimizing to avoid this resulted in a nearly flat rotational transform profile just above $\iota=0.601$. To generate a starting point, we used the method detailed in \cref{subsec:heuristic}.

The aspect ratio of this configuration, calculated in \texttt{VMEC}, is around $A=10$, the mirror ratio $\Delta$ is just below $0.19$, and the maximum flux-surface elongation of this configuration is just above $\varepsilon=5.5$. These values match their respective cut-in values $A_*$, $\Delta_*$, and $\varepsilon_*$ very closely. 

\subsection{Two field-periods}\label{subsec:nfp2}
This optimized configuration has $\Delta=0.22$, $A=10$, and maximum $\varepsilon=6$. Optimizing with no target for $\iota$ resulted for $\iota\in[0.45,0.55]$. While \texttt{SPEC} did not detect any islands in this configuration, we nonetheless continued to optimize this configuration with an explicit penalty for low-order rationals in $\iota$ to find $\iota\in[0.61,0.62]$ to avoid rational surfaces. These are both significant departures from the rotational transform of the configuration used as an initial condition, which sat around $\iota=0.1$, although the shape of the axis looks extremely similar. We discuss this change in \cref{subsec:axismod}.

\subsection{Three field-periods}\label{subsec:nfp3}
This configuration has $\Delta=0.2$, $A=8$, and maximum $\varepsilon=6.2$. The rotational transform $\iota$ lies between 0.75 and 0.8, chosen to avoid low-order rational surfaces. As an initial condition, we used the fully optimized $n_{fp}=2$ configuration, with an increased $n_{fp}$ and decreased number of \texttt{VMEC} modes (\texttt{mpol} = \texttt{ntor} = 2).

\section{Analysis}
\subsection{Confinement}
\begin{figure}
    \centering
    \includegraphics[width=\textwidth]{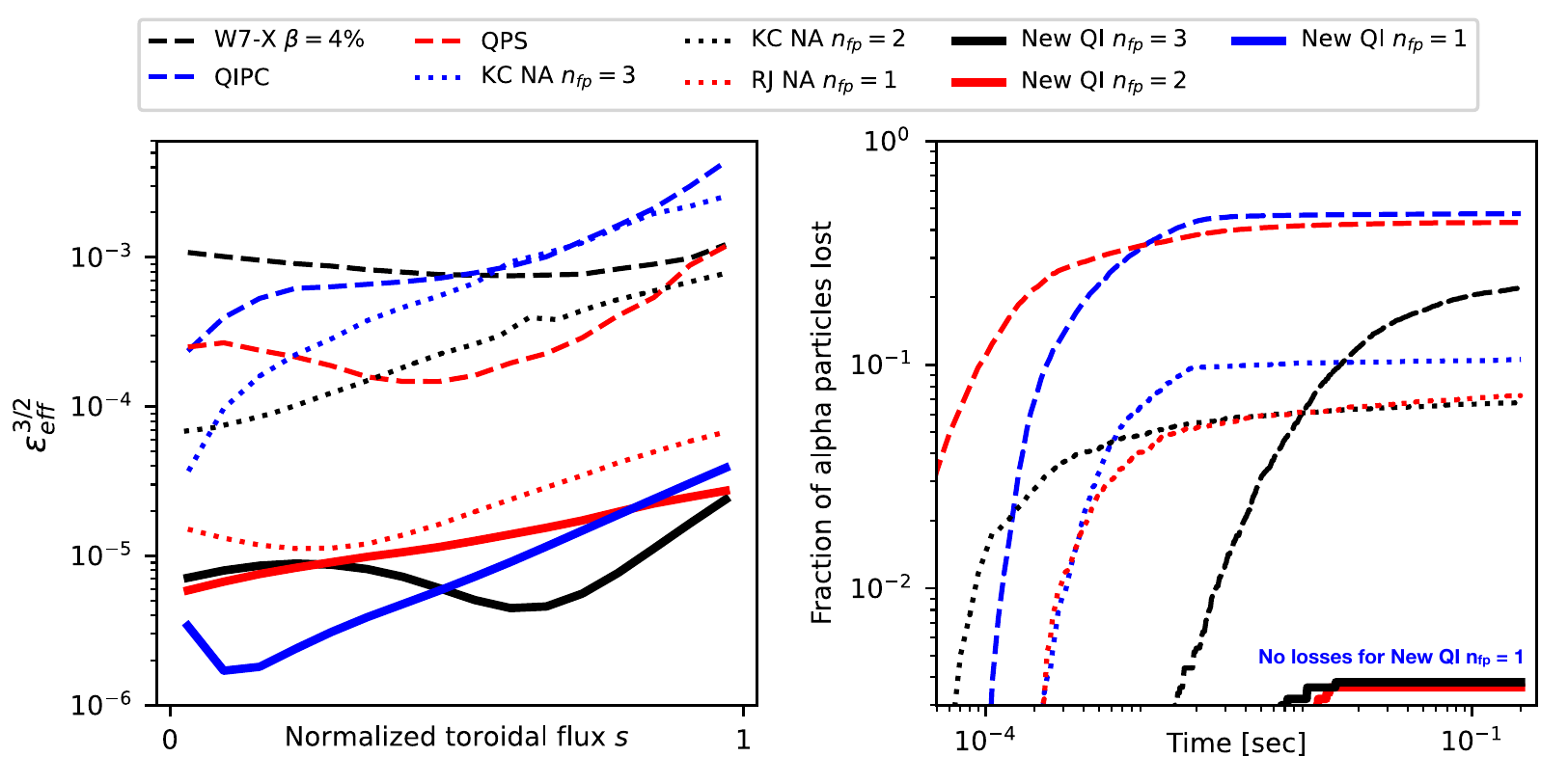}
   \caption{Figures showcasing the confinement properties of these configurations in vacuum, labeled ``New QI'', juxtaposed with the same metrics for other configurations. Dashed lines describe ``legacy'' configurations Wendelstein 7-X \cite{w7x}, QIPC \cite{qipc}, and QPS \cite{qps}, dotted lines describe newer configurations constructed using the near-axis expansion \cite{jorge2022,mata2022}, and thick solid lines are from this work. \textbf{Left:} neoclassical transport coefficient $\varepsilon_\textrm{eff}^{3/2}$. \textbf{Right:} collisionless losses of fusion-generated alpha particles, calculated by \texttt{SIMPLE} \cite{simple}, initialized at $s=0.25$ in an ARIES-CS-scale reactor \cite{aries-cs}.}
    \label{fig:ees_losses_vacc}
    \end{figure}

Compared with most earlier stellarators, all three configurations presented in this work have excellent confinement properties, both in vacuum and with finite pressure. Their vacuum neoclassical transport and fusion-born alpha particle confinement are both shown in \cref{fig:ees_losses_vacc}. To be consistent with \cite{landreman2022}, we scaled these configurations to the size of the ARIES-CS device, i.e., to have an effective minor radius of $1.7$ m (as calculated by \texttt{VMEC}) and field strength $B_{0,0}(s=0)=5.7$ T, and simulated 5000 collisionless alpha particles with energy $3.5$ MeV started isotropically on the $s=0.25$ surface using the \texttt{SIMPLE} code \cite{simple}.

It is interesting to understand how these configurations, and their confinement properties, change when a pressure profile $p(\psi)$ is added to them. In this work, we use a linear pressure profile $p(s) \propto 1 - s$, and use volume ($V$) averaged $\beta$,
\begin{align}
    \beta = \frac{1}{V(s=1)}\int_0^1 \frac{2\mu_0}{B(s)^2} p(s) \frac{\textrm{d}V(s)}{\textrm{d}s} \textrm{d}s,
\end{align}
to describe differences in pressure normalization, where $\mu_0$ is the vacuum permeability constant.

\begin{figure}[ht]
    \begin{center}
        \includegraphics[width=\linewidth]{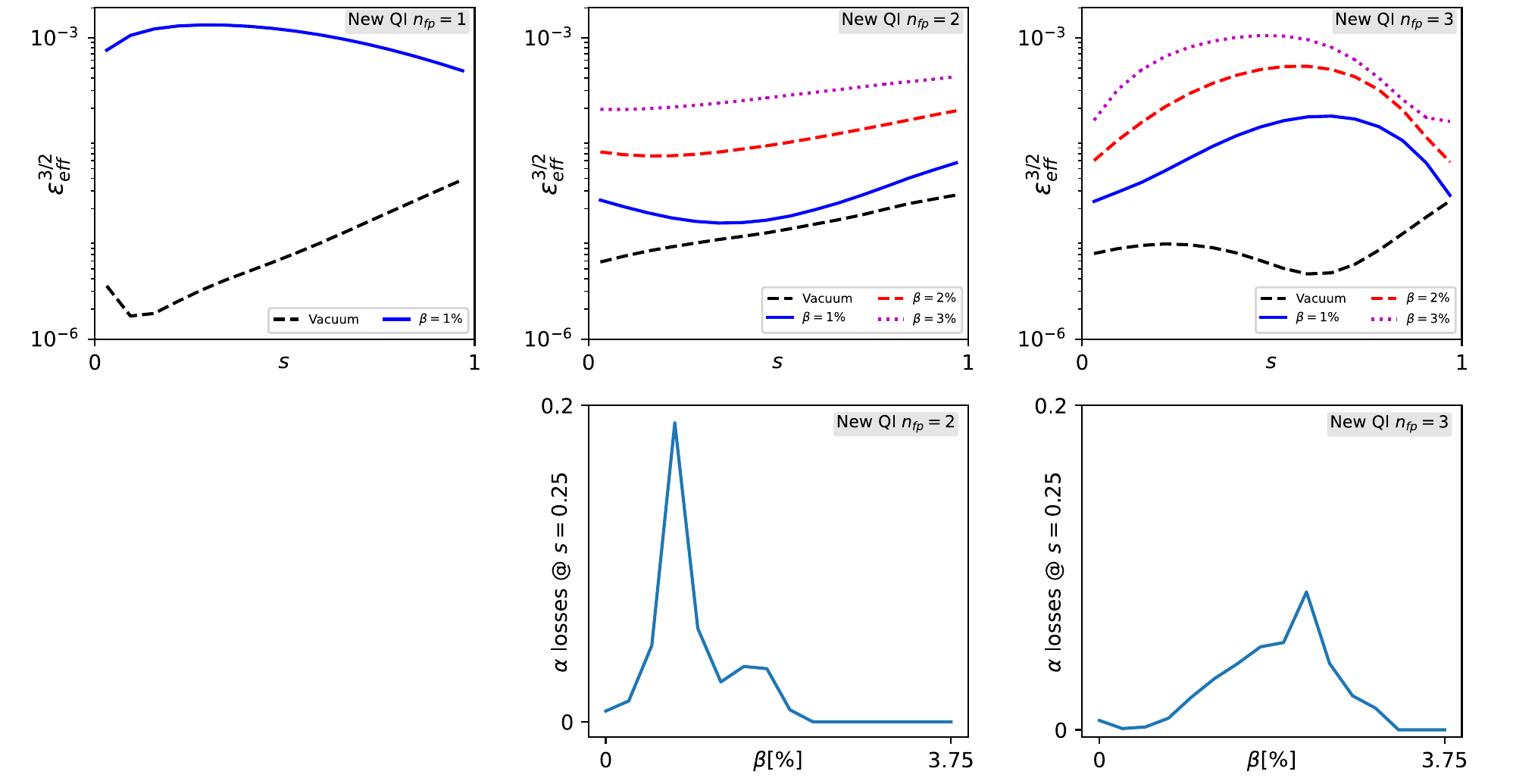}
        \captionsetup{justification=centering}
        \caption{\textbf{Top:} Neoclassical transport coefficient $\varepsilon_\textrm{eff}^{3/2}$ for the three new configurations presented at various $\beta$. \textbf{Bottom:} fraction of 3.5 MeV alpha particle lost, initialized at $s=0.25$, for various values of $\beta$ for the two and three field-period configurations.}
        \label{fig:ees_beta}
    \end{center}
\end{figure}

Because they were optimized in vacuum, we expect neoclassical transport to worsen when more pressure is added (which matches the observed trends in \cref{fig:ees_beta}). For the same reason, $\partial_\alpha\mathcal{J}$ will increase at higher $\beta$, which would lead to more fast-particles to be lost, all else being equal. However, other competing factors create a non-trivial correlation between fast-particle losses and $\beta$. This can be understood by first noting that the plasma is, broadly speaking, diamagnetic, meaning that the magnetic field strength roughly decreases with added plasma pressure, and thus also tends to decrease with radius (in some average sense) when $\textrm{d}p/\textrm{d}s < 0$. Because $\mathcal{J}$ increases when $B$ decreases, we thus expect $\partial_s\mathcal{J}$ to become more negative as plasma pressure builds up. In other words, increased pressure, in general, makes it easier to find maximum-$\mathcal{J}$ configurations, in which $\partial_s\mathcal{J} < 0$. 

Now consider a configuration which is minimum-$\mathcal{J}$ in vacuum (as all known configurations are), meaning that $\mathcal{J}(s_1) < \mathcal{J}(s_2)\,\,\forall\,\,\alpha,\lambda,s_1<s_2$. If enough pressure is added, we expect the configuration to change such that $\mathcal{J}(s_1) > \mathcal{J}(s_2)$. Due to the intermediate value theorem, there must therefore be some ``catastrophic'' $\beta$ at which $\mathcal{J}(s_1) = \mathcal{J}(s_2)$. At this $\beta$, we thus expect more particles to be lost, since particles can drift radially whilst conserving $\mathcal{J}$, even if $\partial_\alpha\mathcal{J}$ is small. 

Despite increased $\partial_\alpha\mathcal{J}$ (see \cref{fig:Jplots}), we still see an improvement in fast particle confinement when $\beta$ is sufficiently large (around 2\% and 3.5\% for the $n_{fp}=2$ and $n_{fp}=3$ configurations respectively), as shown in \cref{fig:ees_beta}. 
This is again caused by the plasma being essentially diamagnetic, such that an increase in $p(s)$ on some flux surface $s$ generally results in decreased $B(s)$. Thus, increased $\beta$ tends to increase $|\partial_s B|$. This creates a ``grad-B'' drift (with a nonzero poloidal component) given by
\begin{align}
    \mathbf{v}_{\nabla B} \propto \frac{\mathbf{B} \times \nabla B}{B^2}.
\end{align}
Thus, increased $\beta$ generally causes particles' poloidal precession to increase. This means that particles that may otherwise drift outwards radially and be lost, can --- at high enough $\beta$ --- poloidally precess quickly enough that their inwards drifts and outwards drifts can negate each other, and the particle can remain confined. For this reason, one can generally expect better fast particle confinement at high $\beta$ \cite{lotz1992,velasco2021}. Physically, changes in poloidal precession are also the cause of the aforementioned catastrophic $\beta$, but it is instructive to think of these two phenomena separately in this case.

To summarize, there are three competing phenomena at play:
\begin{align}
    (\uparrow\beta) & \Longrightarrow
    \left(\uparrow\frac{\partial\mathcal{J}}{\partial\alpha}\right) \Longrightarrow 
    (\uparrow\textrm{ fast particle losses}), \nonumber \\
    & \Longrightarrow
    \left(\downarrow\frac{\partial\mathcal{J}}{\partial s}\right) \Longrightarrow
    (\updownarrow\textrm{ fast particle losses}), \nonumber \\
    & \Longrightarrow
    \left(\uparrow \frac{\mathbf{B}\times\nabla B}{B^2} \right) \Longrightarrow
    (\downarrow\textrm{ fast particle losses}) \nonumber. 
\end{align}
Due to these three factors, the relationship between $\beta$ and fast particle confinement is therefore non-trivial and worth investigating.

\subsection[J properties]{\texorpdfstring{$\mathcal{J}$}{J} properties}
For the above reasons, it is instructive to understand how $\tilde{\mathcal{J}}$ varies with respect to $s$, $\alpha$, and $\lambda$. To extract this information from a \texttt{VMEC} configuration, we begin by finding the \texttt{VMEC} coordinates corresponding to various field lines on a single flux surface using an algorithm used in the \texttt{stella} \cite{stella} code. With these, we linearly interpolate the magnetic field strength along each field line, and use a numerical root-finding algorithm to find bounce points $\phi_1$ and $\phi_2$ at which $B=B_*$ for some $B_\textrm{min}<B_*<B_\textrm{max}$. For a single pair of bounce points along a single field line, we rewrite \cref{eq:J2} in terms of parameters available in $\texttt{VMEC}$:
\begin{equation}
    \tilde{\mathcal{J}}(s,\alpha,B_*) = 
    \int_{\phi_1(s,\alpha,B_*)}^{\phi_2(s,\alpha,B_*)}
    \frac{B\sqrt{1 - \lambda B}}{\mathbf{B}\cdot\nabla\phi}\,\textrm{d}\phi.
\end{equation}
Our algorithm to compute $\tilde{\mathcal{J}}$ takes care to consider which wells along neighbouring field lines are ``connected'' to each other, which allows us to trace field lines over several toroidal turns. It also means that the possibility of transitoning particles is accounted for in our calculation of a single particle's $\tilde{\mathcal{J}}$ along various field lines.

\begin{figure}[ht]
    \begin{center}
        \includegraphics[width=0.92\linewidth]{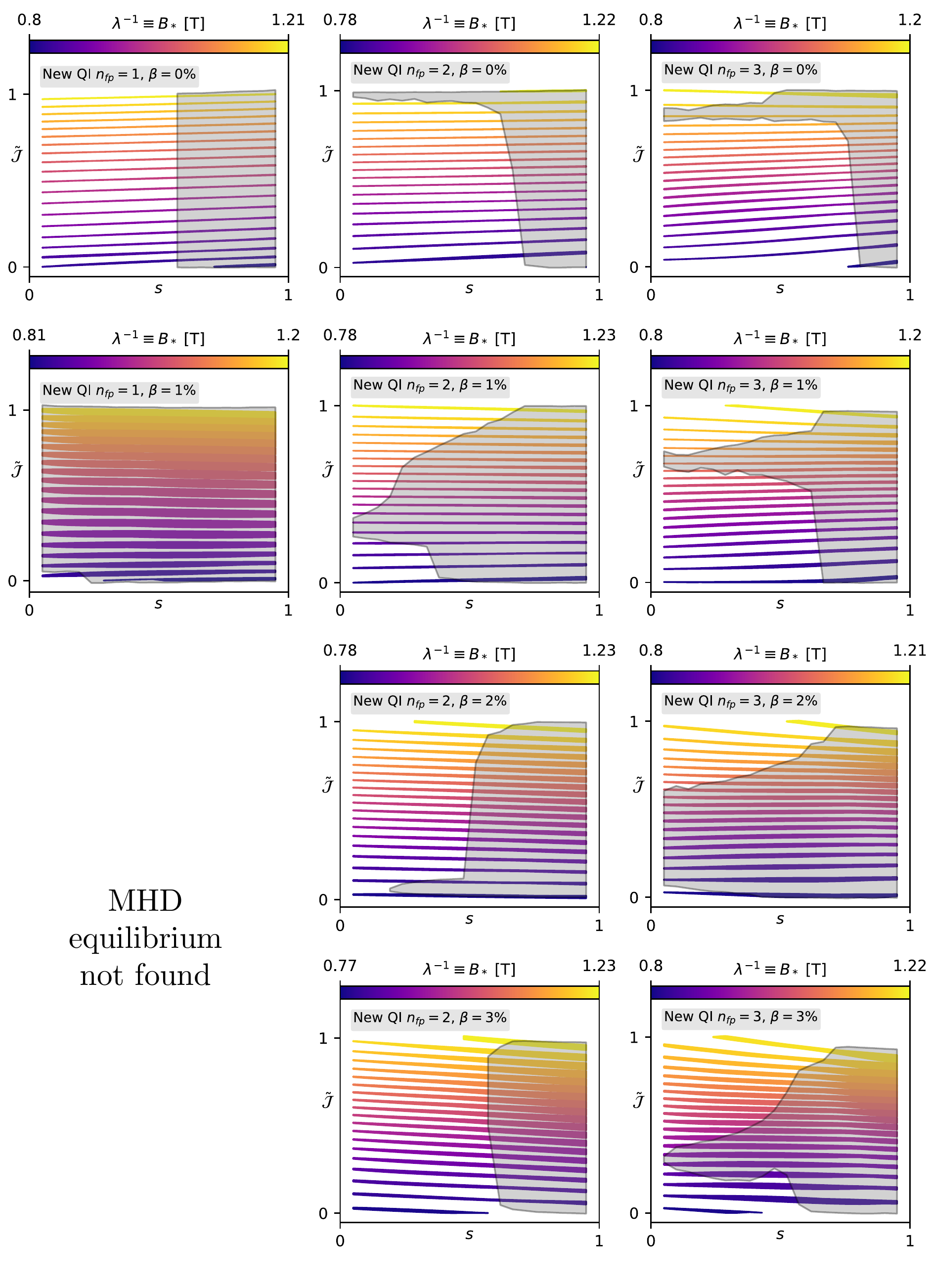}
        \captionsetup{justification=centering}
        \caption{Contours of constant $B_*$ at various $s$ and $\tilde{\mathcal{J}}$ for the three new configurations (left-to-right) at various $\beta$ (top-to-bottom). Important features are (1) the width of these lines, which is bounded by the maximum and minimum $\tilde{\mathcal{J}}$ for each $B_*$ and $s$. Hence, thinner lines means better QI; (2) the slope of these lines $\partial\mathcal{J}/\partial s$, which is negative for maximum-$\mathcal{J}$; (3) the range of values of $\tilde{\mathcal{J}}$ for which particles initialized at various surfaces $s$ are lost, shown by the shaded regions.}
        \label{fig:Jplots}
    \end{center}
\end{figure}

Although $\partial_\alpha\tilde{\mathcal{J}}$ is small for these configurations, it is not exactly zero, and thus a single $B_*$ on a flux surface will be associated with various values of $\tilde{J}$, each corresponding to a different field line. In \cref{fig:Jplots}, we plot $\tilde{\mathcal{J}}$ as a function of $s$, for various bounce field strengths $B_*\equiv1/\lambda$.

From \cref{fig:Jplots}, one can deduce three important properties:

\begin{enumerate}[(1)]
    \item The variation of $\tilde{\mathcal{J}}$ between various field lines on a single flux surface can be seen by the width of the line, which is bounded between $\textrm{max}_\alpha(\tilde{\mathcal{J}}(s,B_*))$ and $\textrm{min}_\alpha(\tilde{\mathcal{J}}(s,B_*))$. In a perfectly omnigenous field, these lines would therefore have zero thickness.
    \item The degree to which a configuration is maximum-$\mathcal{J}$ can be seen by the slope of the lines. In a completely maximum-$\mathcal{J}$ configuration, all lines on these plots would have negative slopes.
    \item Particularly large losses occur when $\partial_s\mathcal{J} = 0$. In this plot, the particles that are lost through this mechanism are those whose $B_* \equiv 1/\lambda$ lines have slopes around zero. The shaded grey (often elephant-shaped) regions show which values of $\tilde{\mathcal{J}}$ have particle losses, which, for these plots, coincide with regions where $\partial_s\mathcal{J} = 0$.
\end{enumerate}

Note in \cref{fig:Jplots} that the $n_{fp}=1$ configuration degrades much more quickly when the pressure is increased to $\beta=1\%$ (as shown by the dramatically increased thickness of the lines of constant $B_*$), and it fails converge in \texttt{VMEC} for $\beta>1\%$. Its magnetic field is, apparently, sensitive to perturbations arising from the diamagnetic and Pfirsch-Schl{\"u}ter (PS) currents arising from the pressure gradient. Configurations with a larger number of field periods are less sensitive to the PS current, the reason for which can be qualitatively understood from the fact that this current closes within one field period of any QI device \citep{helander2014}. In a tokamak, the PS current runs in one direction on the outboard side of the torus and in the other direction on the inboard side. In contrast, it ``turns around'' in each field period of a QI stellarator and never crosses the $B_\textrm{max}$ contour. If the number of periods is large, the current thus has to change direction frequently and therefore does not affect the magnetic geometry very much. In a configuration with only one field-period, the PS current traverses one full toroidal turn around the torus before changing directions, and therefore has a larger effect on the magnetic geometry than if $n_{fp} > 1$. The other two configurations, therefore, are much more resilient to changes in pressure.

Another interesting phenomenon that can be seen from \cref{fig:ees_beta,fig:Jplots} is that there can exist in QI configurations a catastrophic value of $\beta$ at which $\partial_s\mathcal{J}(\lambda) \rightarrow 0$ for a wide range of $\lambda$, resulting in significant fast particle losses. From \cref{fig:ees_beta,fig:Jplots}, we see that this happens at $\beta\simeq1\%$ in the new $n_{fp}=2$ configuration, and at $\beta\simeq2\%$ in the new $n_{fp}=3$ configuration. In a practical stellarator design, one may want to ensure this catastrophic $\beta$ has a small value, so that (1) the field transitions to maximum-$\mathcal{J}$ quickly, and (2) that the increase in alpha particle losses happens well before thermonuclear ignition. 

It is worth mentioning that the size of the grey region is not necessarily proportional to the number of fast particles lost. For instance, it may be the case that all particles are confined, except for a small number of those with $\tilde{\mathcal{J}}\simeq0$, and a small number with $\tilde{\mathcal{J}}\simeq1$. This would result in a grey region that fully covers the plot, despite only a small number of particles being lost.

\subsection{Bootstrap current}\label{subsec:boostrap}
\begin{figure}[ht]
    \centering
    \includegraphics[width=\textwidth]{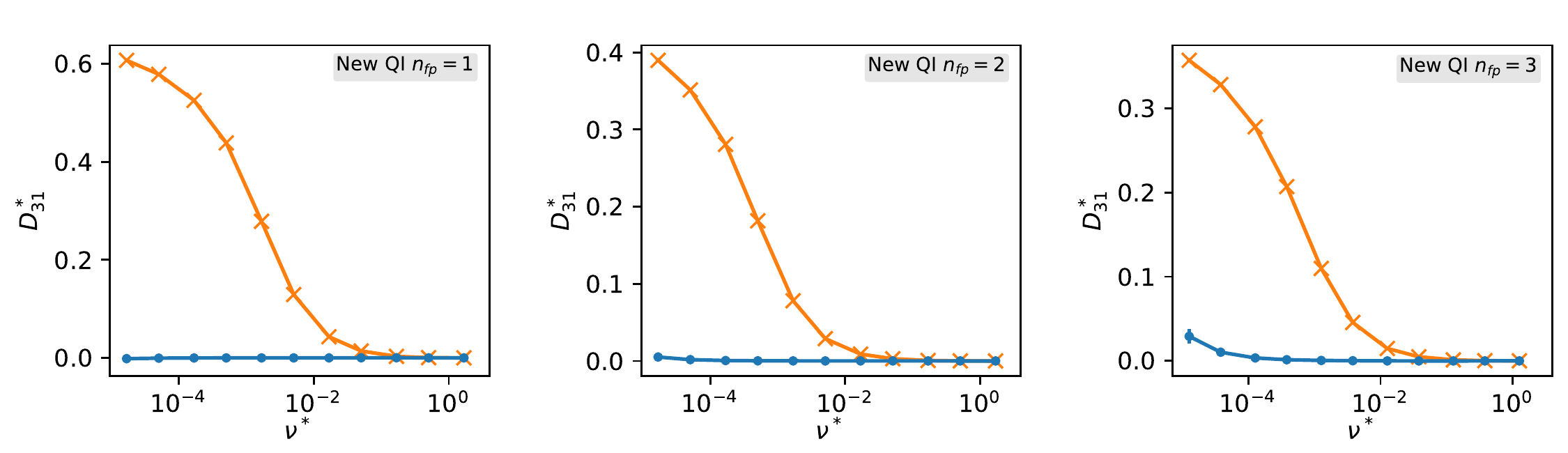}
    \caption{The mono-energetic bootstrap current coefficient $D_{31}^\star$ for the three QI configurations as a function of collisionality $\nu^\star$ (blue) calculated by the DKES code \cite{vanriij1989}, juxtaposed with these coefficients for a tokamak with the same $\iota$, aspect ratio and elongation (orange). These results include error bars where these are large enough to be visible.}
    \label{fig:bootstraps}
\end{figure}

The authors of \cite{helander2009,helander-2011} show that precisely QI configurations have a very small bootstrap current. In this section, we investigate the mono-energetic bootstrap current coefficient $D_{31}^\star$ (as defined in \cite{Beidler_2011}) for the three new QI configurations. Shown in \cref{fig:bootstraps} the coefficient is expectedly very small compared to an equivalent tokamak with the same aspect ratio, elongation and $\iota$ for a wide range of collisionalities $\nu^\star\equiv R_\textrm{maj}\nu/(\iota v)$. Here, $R_\textrm{maj}$ is the stellarator's effective major radius, and $\nu$ is the collision frequency of a particle with velocity $v$.

\subsection[TEMs and available energy]{Trapped-electron modes \& available energy}
In configurations that are both perfectly QI and maximum-$\mathcal{J}$ (meaning $\partial_\psi \mathcal{J} < 0, \forall \, \psi$), it has been shown that density-gradient-driven collisionless trapped-electron modes (TEM) are both linearly and nonlinearly stable. The linear stability criterion, derived from the gyrokinetic equations in \cite{proll2012,helander2013}, reduces to the condition that precessing trapped particles should not be in resonance with a drift wave. 

Following the method of \cite{Proll2014}, this phenomenon can be understood by first using \cref{eq:particle_drifts}, which tells us that the bounce-averaged presessional frequency of trapped particles $\textrm{d}\alpha/\textrm{d}t$ in an omnigenous field is 
\begin{equation}
    \bar{\omega}_\textrm{bnc} = -\frac{1}{Ze\tau_{b}} \partial_\psi\mathcal{J},
\end{equation}
\noindent where $\tau_{b}$ is the particle's bounce time. 

Drift waves are driven by density or temperature perturbations in the plasma and propagate with a frequency proportional to
\begin{equation}
    \omega_\textrm{dia} \propto \frac{T}{e} \frac{\textrm{d}\ln(n)}{\textrm{d}\psi}
\end{equation}
\noindent where $n$ denotes the number density and $T$ the plasma temperature. Because plasma density tends to decrease as a function of $\psi$, it is usually the case that $\textrm{d}\ln(n) / \textrm{d}s < 0$. Hence, we find that
\begin{align}
    \bar{\omega}_\textrm{bnc} \omega_\textrm{dia} \propto -\frac{T}{Ze^2\tau_\textrm{bnc}} \frac{\textrm{d}\ln(n)}{\textrm{d}\psi} \partial_\psi \mathcal{J}, \\
    \textrm{sign}(\bar{\omega}_\textrm{bnc} \omega_\textrm{dia}) = \textrm{sign}\left(\partial_\psi \mathcal{J}\right). \label{eq:sgnww}
\end{align}
TEM-driven turbulence occurs when $\bar{\omega}_\textrm{bnc}$ and $\omega_\textrm{dia}$ are in resonance. This resonance is impossible if \cref{eq:sgnww} is negative, which occurs in a maximum-$\mathcal{J}$ ($\partial_\psi \mathcal{J} < 0$) omnigenous field.\footnote{Note that this stability criterion only holds when the ratio of the electron temperature gradient to the density gradient is in the range $0 \leq \partial_\psi \ln T_e / \partial_\psi \ln n < 2/3$, as shown in \cite{proll2012,helander2013,helander2017available}.} The configurations presented become completely maximum-$\mathcal{J}$ at high $\beta$, and we may expect trapped electron-driven turbulence to decrease as a function of $\beta$ \cite{connor1983effect}.

The relationship between TEM turbulence and maximum-$\mathcal{J}$ configurations was expanded upon in \cite{helander2020available}, where the so-called ``available energy'' (\textit{\AE}) of trapped electrons was calculated. \textit{\AE} measures the maximal amount of thermal energy that may be liberated from the plasma distribution function, subject to some constraints. 

In order to compare the devices fairly, we opt to calculate the fraction of the total thermal energy of the electrons that is available. Using the expression for \textit{\AE} given in \cite{Mackenbach2020} we define this fraction as
\begin{equation}
    \frac{\textrm{\textit{\AE}}}{E_{th}} \equiv \frac{ \int \textrm{\textit{\ae}}(s) \frac{\mathrm{d} V}{\mathrm{d} s} \mathrm{d} s }{\int \frac{3}{2} n(s) T_e(s) \frac{\mathrm{d} V}{\mathrm{d} s} \mathrm{d} s  },
    \label{eq:EsubA}
\end{equation}
where \textit{\ae}$(s)$ is the \textit{\AE} per unit volume. Note that the appropriate length-scale over which the intergrals are to be taken is set by the turbulence correlation length \citep{Mackenbach2020}, which we take to be the poloidal gyroradius given by
\begin{equation}
    \rho_\mathrm{pol} = \frac{\sqrt{2 m T_e(s)}}{\iota Z e B_0 },
\end{equation}
where $B_0$ is the average magnetic field strength on axis. To find the \textit{\AE} per unit volume, one then divides by the correlation volume. In this analysis, we chose $T_e = (1 - s)^4$ and $n = (1 - s)$ as the electron temperature and density profiles respectively, though it should be noted that the observed trends are fairly independent of profile choices.
\par
\begin{figure}[ht]
    \begin{center}
        \includegraphics[width=0.65\linewidth]{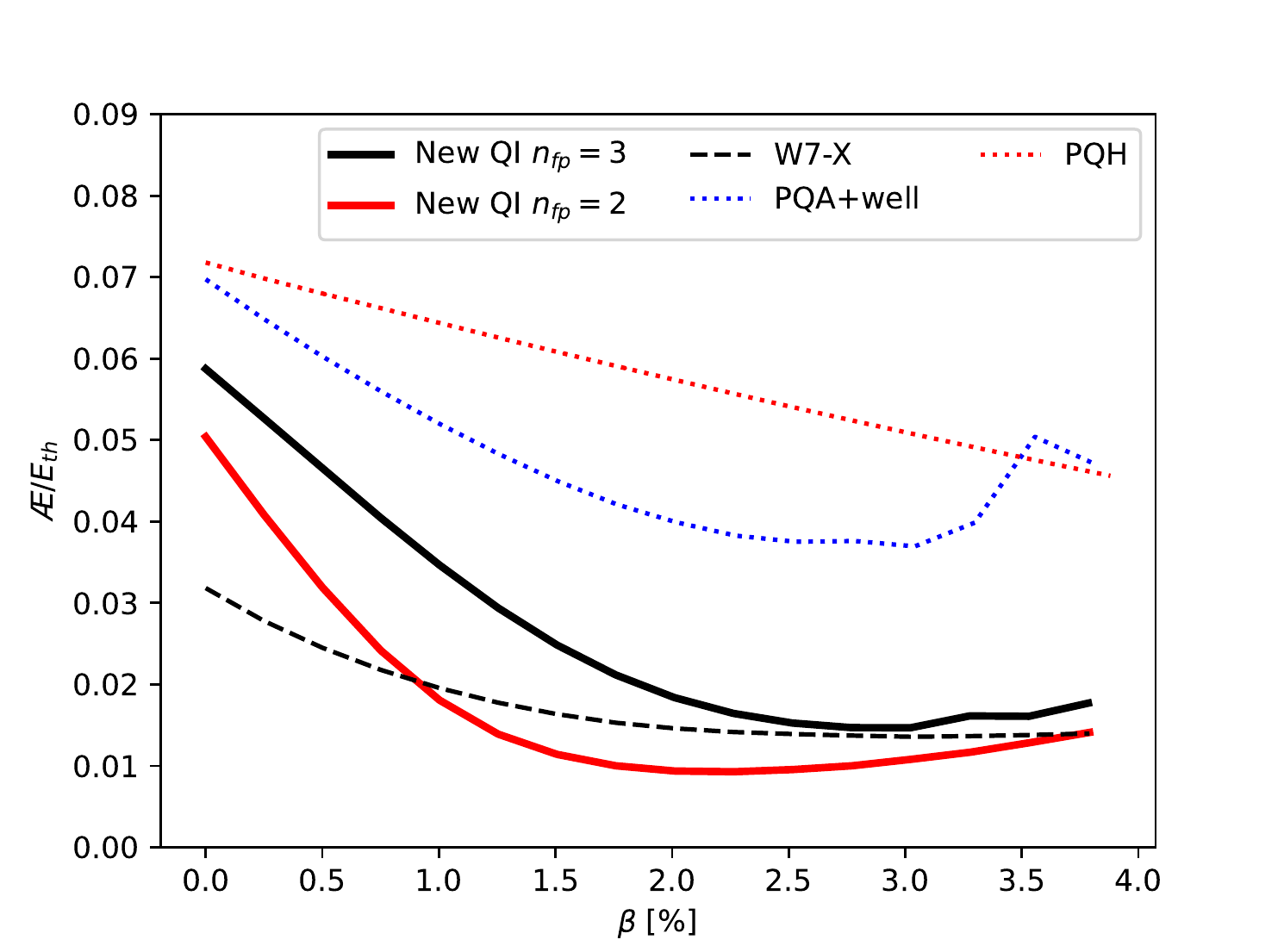}
        \captionsetup{justification=centering}
        \caption{The fraction of thermal energy available for TEM-drive turbulence, \textit{\AE}/$E_{th}$, at various values of $\beta$, for the new configurations presented in this paper (``New QI''), Wendelstein 7-X (W7-X), a precise quasi-axissymmetric configuration with a magnetic well (PQA+well), and a precise quasi-helically symmetric configuration (PQH) \cite{landreman2022}.}
        \label{fig:AEplot}
    \end{center}
\end{figure}

 
The result of this analysis can be seen in \cref{fig:AEplot}, where $\textrm{\textit{\AE}}/E_{th}$ is calculated for the new QI configurations presented, as well as W7-X and the precisely quasisymmetric configurations found in \cite{landreman2022}. 

Several interesting trends can be observed: as expected $\textrm{\textit{\AE}}/E_{th}$ decreases as a function of $\beta$ for small $\beta$, likely as a result of these configurations becoming more maximum-$\mathcal{J}$. Furthermore, at higher $\beta$ the fraction $\textrm{\textit{\AE}}/E_{th}$ tends to increase again due to the loss in omnigenity. Also, the three QI configurations have lower $\textrm{\textit{\AE}}/E_{th}$ than the quasisymmetric configurations. It is worth noting, though, that the new QI configurations presented have comparable (and, at some values of $\beta$, higher) $\textrm{\textit{\AE}}/E_{th}$ than W7-X. We suspect the reason for this is the higher mirror ratios in these new QI configurations, which increases the number of trapped particles, and therefore the total available energy stored in the trapped electrons. This is an interesting result, as it indicates that targeting low $\textrm{\textit{\AE}}/E_{th}$ may be an effective way of limiting the mirror ratio in QI optimizations.

%===========================================================================

\subsection{Axis modifications}\label{subsec:axismod}
Because optimizations with $n_{fp}>1$ were extremely sensitive to initial conditions, it is useful to understand how these optimizations shaped the magnetic axis, such that we can learn to construct better QI configurations using the NAE. In this section, we showcase comparisons between the axis used for the NAE construction, the axis shapes in the optimized \texttt{VMEC} equilibria, and the axis of the optimization's starting point. For brevity, we compare these properties only for the $n_{fp}=2$ configuration presented, as its optimized properties (namely, its rotational transform) deviated most significantly from its initial condition.

\begin{figure}[ht]
    \begin{center}
        \includegraphics[width=\linewidth]{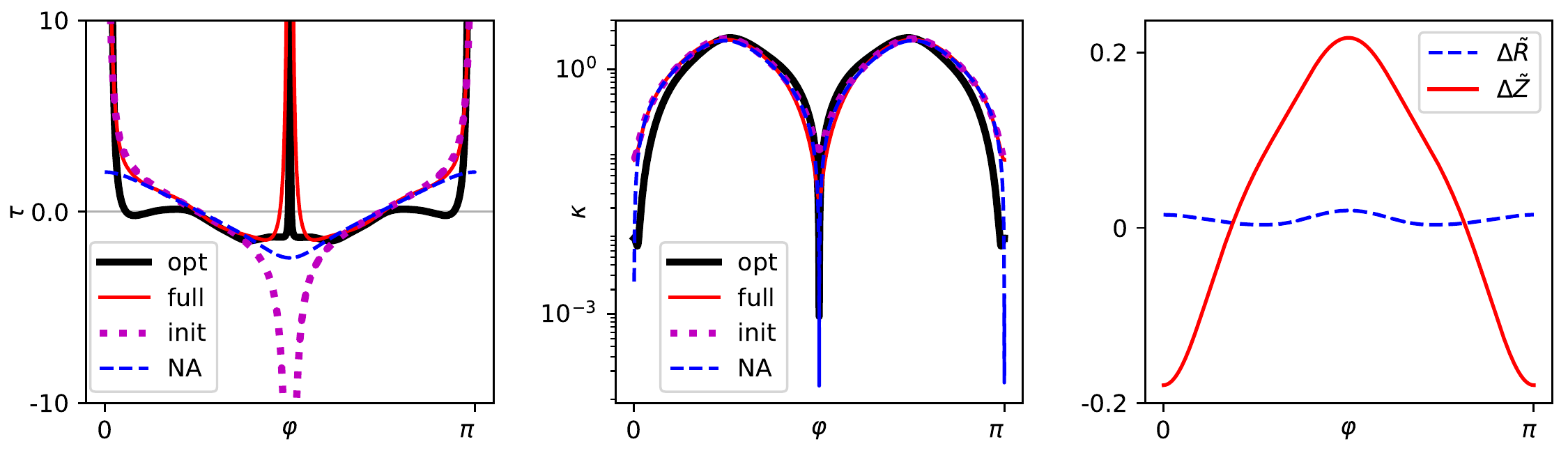}
        \captionsetup{justification=centering}
        \caption{\textbf{Left:} torsion of the optimized magnetic axis (opt), the near-axis construction with a large number of $\texttt{VMEC}$ poloidal and toroidal modes (full), the optimization's initial condition (init), and the axis found using the near-axis expansion before being run through \texttt{VMEC} (NA).
        \textbf{Middle:} the curvature of these same magnetic axes.
        \textbf{Right:} the relative difference between the $R$ and $Z$ coordinates of the optimized configuration's axis and the initial condition's axis.
        }
        \label{fig:axisplots}
    \end{center}
\end{figure}

The shape of a magnetic axis can be understood through either (1) the axes' curvatures $\kappa$ and torsions $\tau$, or (2) the $R$ and $Z$ coordinates of the axes at various toroidal angles. As in \cite{mata2022}, we define curvature and torsion as
\begin{equation}
\begin{gathered}
    \kappa\mathbf{n_a} = \frac{\textrm{d}\mathbf{t_a}}{\textrm{d}l_a},\\
    \tau\mathbf{n_a} = -\frac{\textrm{d}\mathbf{b_a}}{\textrm{d}l_a},
\end{gathered}
\end{equation}
where $\mathbf{t_a}$, $\mathbf{b_a}$, $\mathbf{n_a}$, and $l_a$ are the axis' tangent vector, binormal vector, normal vector, and arclength respectively. We also define the relative $R$ and $Z$ differences between two magnetic axes as
\begin{equation}
\begin{gathered}
    \Delta\tilde{R}(\phi) = \frac{R_{ax}^\textrm{(opt)}(\phi) - R_{ax}^\textrm{(init)}(\phi)}{R_{ax}^\textrm{(opt)}(\phi) + R_{ax}^\textrm{(init)}(\phi)}, \\
    \Delta\tilde{Z}(\phi) = \frac{Z_{ax}^\textrm{(opt)}(\phi) - Z_{ax}^\textrm{(init)}(\phi)}{Z_{ax}^\textrm{(opt)}(\phi) + Z_{ax}^\textrm{(init)}(\phi)}, \\
\end{gathered}
\end{equation}
where (opt) and (init) correspond to the optimized configuration's axis and the initial condition's axis respectively. Plots showcasing these differences are shown in %\cref{fig:AxisComparison,fig:AxisTauComparison}.
\cref{fig:axisplots}.

The presence of zeros of curvature is an intrinsic requirement for QI configurations in the near-axis formalism \cite{Plunk19}. 
Although Fourier mode truncation results in this property being lacking in the configuration used as an initial point for the optimization, the resulting optimized configurations have a curvature function that is approximately zero at points of maxima and minima of $B$ ($\phi=0,\pi/n_{fp}$). 
This indicates that the near-axis solutions, which impose this feature analytically, are optimal configurations in this region of the solution space. The curvature is made small over a wider interval, which should help with the reduction of the first order correction to the magnetic field $B_1$, according to the near-axis theory. 

Another striking feature on the axis shapes of the optimized configurations is the presence of regions where torsion increases sharply. 
This behaviour is expected where curvature approaches zero, but in these cases it is also present in regions of non-zero curvature.
We suspect that this is because, at finite aspect ratio, the configurations constructed using the near-axis expansion do not exactly represent an MHD equilibrium, and thus are only approximately reconstructed when using an equilibrium solver.
As a consequence, the axis shape used for the construction before begin run through \texttt{VMEC} (labelled ``NA'' in \cref{fig:axisplots}) and the one found by \texttt{VMEC} (labelled ``full'' in \cref{fig:axisplots}) are not identical even when a large number of Fourier modes are used in VMEC. This is evident when comparing their curvature and torsion. 
Peaked torsion profiles, like the ones observed in
\cref{fig:axisplots}, are thus commonly observed when using numerical equilibrium solvers. 
According to near-axis theory a sign change is expected in the Frenet frame at locations of zero curvature \citep{mata2022}. 
Numerical approximations of such curves achieve this flip by a rapid ``rotation'' of the axis, explaining the large torsion where the curvature approaches zero.
We also observe an increase in the optimized configuration's integrated torsion (it is more negative) compared to the initial condition, even when ignoring the regions of peaked torsion around $\mathcal{B}_{\mathrm{max}}$ and $\mathcal{B}_{\mathrm{min}}$. This explains the large increase in rotational transform seen in the optimized $n_{fp}=2$ configuration ($\iota\sim0.6$) compared to the rotational transform of the near-axis construction ($\iota\sim0.1$). 

%===========================================================================

\section{Discussion} \label{sec:discussion}
%===========================================================================
We have demonstrated that, using our novel target function, excellent QI quality is achievable in stellarators with reasonable aspect ratios, mirror ratios, and elongations. To our knowledge, the property of quasi-isodynamicity is satisfied to higher accuracy than in any previous publications.  As a result, our configurations have extremely low neoclassical transport, fast-particles simulated in them are well-confined, and their bootstrap currents are vanishingly small. When enough pressure is added, the two- and three-field-period configurations become maximum-$\mathcal{J}$ while maintaining excellent omnigenity, and should thus benefit from TEM stability and little turbulence associated with such modes. 

The three stellarators presented are visually very different from each other, and have different numbers of field-periods, $n_{fp}$, indicating the robustness of this target function. When optimizing configurations with $n_{fp}=1$, we had very good results with every initial condition we tried, but the success of optimizations with $n_{fp}>1$ was quite sensitive to the chosen initial conditions. Future investigation into this phenomenon is warranted, since this trend is consistent with near-axis construction results, which seem to degrade with increased field-periods \cite{jorge2022,mata2022}. These near-axis constructions of this nature seem to work as promising initial conditions for optimization for $n_{fp}>1$.

In this work, we directly penalized mirror ratio and elongation, as excessive values of these quantities are undesirable for fusion reactors. However, it is not clear what exactly the highest allowable value of these parameters should be, and if this is the best way to penalize them. High mirror ratios are undesirable because they imply an inefficient use of magnetic energy (much of it concentrated in small regions) and lead to large forces on the coils that generate the fields. A target function that directly accounts for these parameters, or simply penalizing $\varepsilon_\textrm{eff}$, may be preferable alternatives. High flux-surface elongations are undesirable because they increase the plasma's area-to-volume ratio. Further, they also tend to increase flux-surface compression which increases instability growth rates \cite{helander2021,stroteich_xanthopoulos_plunk_schneider_2022}. High elongation also complicates coil optimization. More carefully chosen targets which don't rely on arbitrary cut-in values may be better suited to future stellarator designs.

Our $n_{fp}=1$ configuration has almost zero shear. The other two have some positive shear, but still very little, which is detrimental to MHD stability.  More troublingly, if left unchecked, all optimizations attempted would converge on rotational transform profiles centered on low-order rationals. Thankfully, these configurations seemed to be fairly robust to changes in $\iota$, and so these low-order rationals were fairly easy to avoid with additional target functions, even after optimizations had already converged reasonably well.

Applications and extensions of this approach to optimization are boundless. Including both MHD and gyrokinetic stability metrics in the target function will be necessary when designing a viable fusion reactor. Finding a set of coils with reasonable properties would also be a valuable next step. The optimization approach in this paper also seems to be sensitive to initial conditions: we were unable to find any good configurations $n_{fp}>1$, except for those initialized from a solution found using near-axis expansion. However, not all near-axis solutions were effective initial conditions. Hence, a more robust and reliable way to generate near-axis solutions as starting points for optimization would be a valuable asset.

%===========================================================================
\section*{Acknowledgements}
%===========================================================================
We are grateful to Christopher Albert, Paul Costello, Michael Drevlak, Albert Johansson, Gregor Pechstein, and Richard Nies for useful conversations. We additionally thank Carolin N{\"u}hrenberg and Donald Spong, who provided the other QI configurations shown. Support from the SIMSOPT team is also gratefully acknowledged.
Optimizations and simulations were performed with the Cobra and Raven HPCs (Garching) respectively. 
A.G., M.L., and K.C.M are supported by a grant from the Simons Foundation (560651). 
R. J. is supported by the Portuguese FCT - Fundação para a Ciência e Tecnologia, under grant 2021.02213.CEECIND and by a grant by Alexander-von-Humboldt-Stiftung, Bonn, Germany, through a research fellowship. 
This publication is part of the project ``Shaping turbulence – building a framework for turbulence optimisation of fusion reactors'', with Project No. OCENW.KLEIN.013 of the research program ``NWO Open Competition Domain Science'' which is financed by the Dutch Research Council (NWO).
This work has been carried out with in the framework of the EUROfusion Consortium, funded by the European Union via the Euratom Research and Training Programme (Grant Agreement No. 101052200 - EUROfusion). Views and opinions expressed are those of the author(s) only and do not necessarily reflect those of the European Union or the European Commission. Neither the European Union nor the European Commission can be held responsible for them.

% Hack added to remove bibliography from TOC
\let\oldaddcontentsline\addcontentsline
\renewcommand{\addcontentsline}[3]{}

\bibliography{mybibliography}

\let\addcontentsline\oldaddcontentsline
\end{document}